\documentclass[twocolumn]{aastex61}

\newcommand{\vect}[1]{\boldsymbol{#1}}
\renewcommand{\thesubsection}{\arabic{subsection}}
\usepackage{makecell}
\usepackage{hyperref}
\usepackage{verbatim}
\usepackage{pbox}
\usepackage{physics}

\newcommand{\mpe}{Max Planck Institute for extraterrestrial Physics, Giessenbachstr., 85748 Garching, Germany}
\newcommand{\lesia}{LESIA, Observatoire de Paris, PSL Research University, CNRS, Sorbonne Universit\'es, UPMC Univ. Paris 06, Univ. Paris Diderot, Sorbonne Paris Cit\'e} 
\newcommand{\mpa}{Max-Planck-Institut f\"ur Astronomie, K\"onigstuhl 17, 69117 Heidelberg, Germany}
\newcommand{\koln}{1. Physikalisches Institut, Universit\"at zu K\"oln, Z\"ulpicher Str. 77, 50937 K\"oln, Germany}
\newcommand{\ipag}{Univ. Grenoble Alpes, CNRS, IPAG, F-38000 Grenoble, France}
\newcommand{\portugal}{Centro Multidisciplinar de Astrof\'isica, CENTRA (SIM), Lisbon and Oporto, Portugal}
\newcommand{\eso}{European Southern Observatory, Karl-Schwarzschild-Str. 2, 85748 Garching, Germany}
\newcommand{\esosantiago}{European Organisation for Astronomical Research in the Southern Hemisphere, Casilla 19001, Santiago 19, Chile}
\newcommand{\geneve}{Observatoire de Gen\`eve, Universit\'e de Gen\`eve, 51 ch. des Maillettes, 1290 Versoix, Switzerland}

\newcommand{\chile}{Unidad Mixta Internacional Franco-Chilena de Astronom\'ia (CNRS UMI 3386), Departamento de Astronom\'ia, Universidad de Chile, Camino El Observatorio 1515, Las Condes, Santiago, Chile}
\newcommand{\bonn}{Max-Planck-Institute for Radio Astronomy, Auf dem H\"ugel 69, 53121 Bonn, Germany}

\newcommand{\berkeley}{Department of Physics, Le Conte Hall, University of California, Berkeley, CA 94720, USA}
\newcommand{\dublin}{Dublin Institute for Advanced Studies, 31 Fitzwilliam Place, D02 XF86 Dublin, Ireland}

\begin{document}

\title{Sub-milliarcsecond Optical Interferometry of the HMXB BP Cru
  with VLTI/GRAVITY}

\correspondingauthor{I.~Waisberg} 
\email{idelw@mpe.mpg.de}

\collaboration{GRAVITY collaboration}
\altaffiliation{GRAVITY is developed in a collaboration
    by the Max Planck Institute for Extraterrestrial Physics, LESIA of Paris Observatory and IPAG
of Université Grenoble Alpes / CNRS, the Max Planck Institute for Astronomy,
the University of Cologne, the Centro Multidisciplinar de Astrof\'isica
Lisbon and Porto, and the European Southern Observatory}

\author{I.~Waisberg}
\affiliation{\mpe}
\author{J.~Dexter}
\affiliation{\mpe}
\author{O.~Pfuhl}
\affiliation{\mpe}

\author{R.~Abuter}
\affiliation{\eso}
\author{A.~Amorim}
\affiliation{\portugal}
\author{N.~Anugu}
\affiliation{\portugal}
\author{J.P.~Berger}
\affiliation{\eso} 
\author{N.~Blind}
\affiliation{\geneve}  
\author{H.~Bonnet}
\affiliation{\eso} 
\author{W.~Brandner}
\affiliation{\mpa} 
\author{A.~Buron} 
\affiliation{\mpe} 
\author{Y.~Cl\'enet}
\affiliation{\lesia}
\author{W.~de~Wit}
\affiliation{\esosantiago}
\author{C.~Deen}
\affiliation{\mpe}
\author{F.~Delplancke-Str\"obele} 
\affiliation{\eso} 
\author{R.~Dembet}
\affiliation{\lesia} 
\author{G.~Duvert}
\affiliation{\ipag} 
\author{A.~Eckart}
\affiliation{\koln} 
\affiliation{\bonn} 
\author{F.~Eisenhauer}
\affiliation{\mpe}
\author{P.~F\'edou}
\affiliation{\lesia}
\author{G.~Finger}
\affiliation{\eso}
\author{P.~Garcia}
\affiliation{\portugal}
\author{R.~Garcia Lopez}
\affiliation{\mpa} 
\affiliation{\dublin}
\author{E.~Gendron}
\affiliation{\lesia}
\author{R.~Genzel} 
\affiliation{\mpe} 
\affiliation{\berkeley} 
\author{S.~Gillessen}
\affiliation{\mpe}
\author{X.~Haubois}
\affiliation{\esosantiago} 
\author{M.~Haug} 
\affiliation{\mpe} 
\affiliation{\eso} 
\author{F.~Haussmann} 
\affiliation{\mpe}
\author{Th.~Henning} 
\affiliation{\mpa} 
\author{S.~Hippler}
\affiliation{\mpa} 
\author{M.~Horrobin}
\affiliation{\koln} 
\author{Z.~Hubert}
\affiliation{\lesia}
\affiliation{\mpa}
\author{L.~Jochum}
\affiliation{\eso} 
\author{L.~Jocou} 
\affiliation{\ipag} 
\author{P.~Kervella}
\affiliation{\lesia}
\affiliation{\chile}
\author{Y.~Kok}
\affiliation{\mpe} 
\author{M.~Kulas}
\affiliation{\mpa}
\author{S.~Lacour}
\affiliation{\lesia}
\author{V.~Lapeyr\`ere}
\affiliation{\lesia} 
\author{J.-B.~Le~Bouquin}
\affiliation{\ipag} 
\author{P.~L\'ena}
\affiliation{\lesia}
\author{M.~Lippa} 
\affiliation{\mpe} 
\author{A.~M\'erand}
\affiliation{\eso}
\author{E.~M\"uller}
\affiliation{\mpa} 
\affiliation{\eso}
\author{T.~Ott}
\affiliation{\mpe}
\author{L.~Pallanca}
\affiliation{\esosantiago} 
\author{J.~Panduro}
\affiliation{\mpa}
\author{T.~Paumard}
\affiliation{\lesia}
\author{K.~Perraut}
\affiliation{\ipag} 
\author{G.~Perrin}
\affiliation{\lesia} 
\author{S.~Rabien}
\affiliation{\mpe} 
\author{A.~Ram\'irez}
\affiliation{\esosantiago}  
\author{J.~Ramos}
\affiliation{\mpa} 
\author{C.~Rau}
\affiliation{\mpe} 
\author{R.-R.~Rohloff}
\affiliation{\mpa}
\author{G.~Rousset}
\affiliation{\lesia}
\author{J.~Sanchez-Bermudez}
\affiliation{\mpa} 
\author{S.~Scheithauer}
\affiliation{\mpa}
\author{M.~Sch\"oller}
\affiliation{\eso}
\author{C.~Straubmeier}
\affiliation{\koln} 
\author{E.~Sturm}
\affiliation{\mpe} 
\author{F.~Vincent}
\affiliation{\lesia}
\author{I.~Wank}
\affiliation{\koln} 
\author{E.~Wieprecht}
\affiliation{\mpe} 
\author{M.~Wiest}
\affiliation{\koln}
\author{E.~Wiezorrek}
\affiliation{\mpe}
\author{M.~Wittkowski}
\affiliation{\eso}
\author{J.~Woillez}
\affiliation{\eso}
\author{S.~Yazici}
\affiliation{\mpe} 
\affiliation{\koln} 

\shorttitle{BP Cru with VLTI/GRAVITY}
\shortauthors{Waisberg et al.}

\begin{abstract}
We observe the HMXB BP Cru using interferometry in the near-infrared K band with VLTI/GRAVITY. 
Continuum visibilities are at most partially resolved, consistent with the predicted size of the hypergiant. Differential visibility 
amplitude ($\Delta |V| \sim 5\%$) and phase ($\Delta \phi \sim 2 \degr$) signatures are observed across the 
HeI $2.059 \mu$m and Br$\gamma$ lines, the latter seen strongly in emission, unusual for the donor star's spectral type. For a baseline $B \sim 100$\,m, the differential phase RMS $\sim 0.2 \degr$ corresponds to an astrometric precision of 
$\sim 2 \mu$as. A model-independent analysis in the marginally resolved limit of interferometry reveals asymmetric 
and extended emission with a strong wavelength dependence. 
We propose geometric models based on an extended and distorted wind and/or a high 
density gas stream, which has long been predicted to be present in this system. The observations show that 
optical interferometry is now able to resolve HMXBs at the spatial scale at which accretion takes place, and 
therefore probe the effects of the gravitational and radiation fields of the compact object on its environment. 

\end{abstract}


\section{INTRODUCTION}
\label{introduction}

X-ray binaries are usually divided into two classes: high-mass (HMXB), in which the compact object is fed by a strong wind/disk 
from a massive OB/Be companion, and low-mass (LMXB), in which accretion happens through Roche lobe overflow from a 
low-mass star, leading to the formation of an accretion disk around the compact object. In both cases, the 
compact object can be a white dwarf, neutron star or a black hole.  

The small scale of such systems, typically with semi-major axis $a < 1$\,mas, means that they are below the imaging resolution 
even of the largest optical/near-infrared interferometers. Therefore, information about the accretion process in these systems and the interaction between the compact object's X-ray output and the stellar environment have so far been restricted to X-ray or 
optical photometry and spectroscopy, from which spatial information are then inferred. 

However, spectral differential interferometry can provide direct
spatial information about such systems on scales as small as $\sim 1-10$ $\mu$as.  There are, however, several challenges. First of all, 
interferometry requires a bright enough object for fringe tracking due to the very 
short atmospheric coherence time that degrades the interferometric signals. For the typical optical/near-infrared 
interferometers working in the V, K or H band, this means that nearly all LMXBs and the great majority of HMXBs 
cannot be observed interferometrically. 

GRAVITY (\citealt{Eisenhauer11}, GRAVITY Collaboration 2017, submitted), the four-telescope beam combiner working at the Very Large Telescope 
Interferometer (VLTI) and which operates in the K band, has made it possible to observe fainter objects and to 
achieve very small differential visibility errors, mainly driven by an improved fringe tracking system, which allows 
for longer coherent integration times, as well as the overall stability of the instrument contributed by its many 
subsystems. In the case of GRAVITY, fringe tracking limits are $K \lesssim 7$ and 
$K \lesssim 10$ for the Auxiliary Telescopes (ATs) and Unit Telescopes (UTs) at VLTI, respectively, which 
means that there are only a handful of Galactic targets that are doable \citep{Liu06, Walter15}. We note that dual-field interferometers such 
as GRAVITY could potentially overcome this difficulty by fringe tracking on a nearby bright reference source, which would 
allow the magnitude limits to be substantially improved, but the small FOV ($2-4$ arcseconds) 
means that such a case is unlikely. 

The only published past observations of a HMXB with an optical interferometer were of Vela X-1 \citep{Choquet14} and 
CI Cam \citep[and references therein]{Thureau09}. The former was 
observed with VLTI/AMBER in the K band and VLTI/PIONIER in the H band. It contains a supergiant O star 
emitting a strong stellar wind and a massive slowly rotating pulsar. Resolved structures of radius $\sim 8 \pm 3 R_*$ 
and $\sim 2 \pm 1 R_*$ were inferred from K and H band continuum visibilities, respectively. Two different interpretations 
were proposed: the resolved structure could be a stellar wind with a strong temperature gradient that deviates significantly 
from a black body at thermal equilibrium, or the resolved structure in the K band was a diffuse shell not present at the time 
of the H band observations, which would then correspond to either the stellar wind or the photosphere. Even though spectral 
lines from HI and HeI were observed in the high resolution K band spectrum, no differential visibility signatures were detected 
beyond the noise level, and therefore the application of differential spectral interferometry was not possible. CI Cam was observed 
with PTI in the K band and with IOTA in the K and H bands. The system is a B(e) X-ray binary and the nature of the compact object is unknown. The interferometric 
observations were able to resolve extended, hot emission from a ring-shaped circumstellar dust envelope of major axis $\sim 8 \text{ mas}$. However, no clear evidence 
for the compact companion was found and the low resolution did not allow the usage of differential spectral interferometry. 

BP Cru is among the brightest HMXBs in the K band ($K=5.7$). It is also one of the canonical wind-accreting 
HMXBs; it has, however, several unique properties, some of which are listed in Table \ref{table:BP Cru Properties}. 
Together with Vela X-1, it contains one of the most massive pulsars known (GX 301-2). Although with a typical magnetic 
field strength of a young neutron star, the pulsar also has one of the
longest spin periods known. The donor star, Wray 977, is a rare
hypergiant of B1Ia+ classification. There are only a handful others in the Galaxy 
\citep{Clark12}, and it is the only one known to be in a binary system. Furthermore, it has one of the most eccentric orbits
among HMXBs. With the goal of studying the inner regions of this system, we have conducted interferometric observations of BP Cru during the commissioning stage of VLTI/GRAVITY in May 2016. This paper reports on these observations. 

\begin{table}[ht]
\centering
\caption{\label{table:BP Cru Properties} Properties of BP Cru / Wray 977 / GX 301-2}
\begin{tabular}{c|c|c|c}
\hline \hline
Parameter & \shortstack{Symbol/\\Unit} & Value & Reference \\ \hline 
\hline 
& BP Cru & & \\ \hline
distance & $d$ (kpc) & $\approx3$ & (1) \\ \hline
orbital period & $P_{orb}$ (days) & \shortstack{$41.498$ \\  $(\pm 0.002)$} & (2) \\ \hline
eccentricity & $e$ & \shortstack{$0.462$ \\ $(\pm 0.014$)} & (2) \\ \hline
\shortstack{binary \\ inclination} & $i$ (deg) & \shortstack{$60$ \\  $(\pm 10$)} & (1) \\ \hline 
\shortstack{mean X-ray \\ luminosity} & $\langle L_X \rangle (\text{ergs/s})$ & $7 \times 10^{36}$ & (1) \\ \hline
\shortstack{maximum X-ray \\ luminosity} & $L_X^{max} (\text{ergs/s})$ & $4 \times 10^{37}$ & (1) \\ \hline
\hline 
& \shortstack{Wray 977 \\ (B1Ia+)} & & \\ \hline
mass & $M_* (M_{\odot})$ & $39-68$ & (1)  \\ \hline
radius & $R_* (R_{\odot})$ & $62^a$ & (1) \\ \hline
\shortstack{photosphere \\ radius} & $R_{2/3} (R_{\odot})$ & $70^b$ & (1) \\ \hline
\shortstack{bolometric \\ luminosity} & $L_* (L_{\odot})$ & $5 \times 10^5$ & (1) \\ \hline
\shortstack{effective \\ temperature} & $T_{eff} (K)$ & \shortstack{$18100^b$ \\  $(\pm 500$)} & (1) \\ \hline
mass-loss rate & $\dot{M} (M_{\odot}/\text{yr})$ & $10^{-5}$ & (1) \\ \hline
\shortstack{wind \\ terminal velocity} & $v_{\infty} (\text{km/s})$ & $305$ & (1) \\ \hline
\shortstack{speed below \\ sonic point} & $v_{2/3} (\text{km/s})$ & $4.40$ & (1) \\ \hline
\shortstack{volume \\ filling factor} & $f$ & $1.0$ & (1) \\ \hline
\shortstack{rotational \\ velocity} & $v \sin i$ (km/s) & $50 \pm 10$ & (1) \\ \hline
\shortstack{radial velocity \\ amplitude} & $K_*$ (km/s) & $10 \pm 3$ & (1) \\ \hline
\hline
& GX 301-2 & & \\ \hline
\shortstack{projected \\ semi-major axis} & $a_X \sin i$ (lt-s) & $368.3 \pm 3.7$ & (2) \\ \hline
\shortstack{radial velocity \\ amplitude} & $K_X$ (km/s) & $218.3 \pm 3.3$ & (2) \\ \hline
mass (lower limit) & $M (M_{\odot})$ & $1.85 \pm 0.6$ & (1) \\ \hline
spin period & $P_{spin} (s)$ & $696$ &  (3) \\ \hline
\shortstack{surface \\ magnetic field} & $B (G)$ & $4 \times 10^{12}$ & (3) \\ \hline
\end{tabular}
\tablenotetext{a}{At Rosseland optical depth $\tau \sim 30$.}
\tablenotetext{b}{At Rosseland optical depth $\tau = 2/3$.}
\tablenotetext{}{References: (1) \cite{Kaper06} (2) \cite{Koh97} (3) \cite{Kreykenbohm04}}
\end{table}

We summarize the relevant background about this system that will guide us in the 
interpretation of the interferometric results (Section 2). Section 3 summarizes the observations and the most important aspects of the data reduction. Section 4 presents the analysis of the K band spectrum. Section 5 presents the interferometric results, which are then discussed and fit to physically inspired geometrical models in Section 6. Section 7 presents complementary data that hints at the future work for this project. Finally, Section 8 summarizes the main results. 

\section{The Effects of the Compact Object on the Surrounding Stellar Environment} 

In this section, we summarize relevant information known about BP Cru that will guide the interpretation of the interferometric results. 
In BP Cru, the pulsar is embedded in the dense stellar wind of Wray 977 and its gravitational and radiation fields are expected to 
substantially influence the surrounding stellar environment. We note that at the orbital phase of observation ($\phi \sim 0.21$ using orbital parameters from \cite{Koh97}), the compact object 
was at a distance $\sim 210 R_{\odot}$ from the donor star's center (the minimum distance at periastron is $\sim 100 R_{\odot}$). 

\subsection{The Accretion Mechanism and the Gravitational Influence of the Pulsar}

As in other HMXBs, the X-ray output of BP Cru is explained through the capture of the strong stellar wind of a 
supergiant companion by the compact object \citep{bondihoyle1944}. X-ray light curves and column densities for many of these
systems, on the other hand, have found evidence of more complex mechanisms, with a spherically symmetric wind accretion model unable to 
explain the data successfully. 
 
 \cite{Stevens88} studied the gravitational effects of the compact object along an eccentric orbit, and 
 found that the wind mass-loss rate is substantially enhanced within a small angle around the line-of-centers, resulting in a higher accretion rate that could 
 explain the X-ray outburst intensities better than a spherically symmetric wind accretion model. This inspired accretion models which included, in 
 addition to the spherical wind, a tidal stream of gas of enhanced density that trails the compact object along its orbit and is responsible for most of 
 the accretion rate. In the case of BP Cru, such models better explain its X-ray emission and column density as a function
 of orbital phase than purely spherical wind models \citep{Haberl91, Leahy91, Leahy02}. In particular, the presence of a strong X-ray outburst slightly
 before periastron, as well as a smaller peak near apastron, could be explained by the pulsar moving through the dense gas stream two times per 
 orbital period. Studies of the X-ray hardness ratio along the orbit are also in rough agreement with such a model \citep{Evangelista10}. Moreover, an increase in column density during superior conjunction points to a stream of enhanced density trailing the X-ray source. 
 The most recent analysis by \cite{Leahy08} found a density enhancement in the stream of $\sim 20 \times$ compared to the wind, 
 resulting in a mass loss rate in the stream $\sim 2.5 \times$ higher than the wind. In this scenario, such a gas stream would then dominate not only the accretion
 process, but also the mass loss itself. For BP Cru in particular, the high eccentricity, which implies that the pulsar's distance from the massive star varies by a 
 factor of $\frac{1+e}{1-e} \sim 2.7$ (the same holding for its speed), can lead to complex stream shapes. \cite{Kaper06} notes that tidal interaction is expected during periastron passage, and
also finds evidence for variations in the emission and absorption parts of the optical P-Cygni lines H$\beta$ and HeI $5876 A$; in particular, a blue-shifted absorption component is seen at all orbital phases. This could be evidence for the presence of a large scale gas stream in the system, both in the orbital plane as well as in the direction perpendicular to it. 

 Models invoking a circumstellar disk around the supergiant star and inclined with respect to the binary plane have also been proposed as an accretion mechanism \citep{Pravdo95}. However, they have found less success than the stream models to explain the X-ray light curve \citep{Leahy02}. Furthermore, there is no evidence of a circumstellar disk in the optical spectrum \citep{Kaper06}.
 
We note that the X-ray light curve of BP Cru is quite stable, with no clear distinction between low/hard and high/soft states typical of systems containing accretion disks. 
 However, \cite{Koh97} reports on two rapid spin-up episodes of the pulsar lasting for about 30 days, and suggests that this may point to the formation of transient 
 accretion disks following a period of increased accretion rate. Furthermore, the recent, first radio detection from BP Cru suggests a variable component in addition to 
 a baseline component arising from Wray 977's wind, and possibly associated with a weak and transient jet \citep{Pestalozzi09}. 
 
\subsection{The Radiation Influence of the Pulsar}

The X-ray emission of the pulsar is expected to influence the surrounding stellar environment, mainly through radiation pressure, X-ray heating and photoionization. 
In hot stars, the wind is accelerated by scattering from photons absorbed in line transitions \citep[CAK model,][]{Castor75}. The ionization of the wind results in a cut off in the wind acceleration, leading to an increase in the wind density that has been evoked to explain the increase in 
accretion rates in systems that undergo transient behavior. At very
high X-ray illumination that suppresses radiative cooling, X-ray
heating can lead to thermally-driven winds \citep{Blondin94}.

\cite{Haberl91} and \cite{Islam14} found evidence for X-ray ionization
of the wind when BP Cru was in outburst near periastron from a low energy excess $\lesssim 3$ keV in the 
 X-ray spectrum. Variations in the X-ray light curve mean brightness between different orbital periods could also point to X-ray irradiation effects \citep{Leahy08}. Finally, we note that recently, about 
 two months before the observations reported in this paper, an unusual and extremely bright X-ray outburst was reported with \textit{Swift} with evidence for strong ionization of the surrounding environment \citep{Fuerst16}. 
 
 In summary, there is ample evidence that the pulsar is closely interacting with the stellar environment in BP Cru. Recent 3D hydrodynamical simulations 
 to study simultaneously the gravitational and radiation effects of the compact object on the stellar wind of HMXBs support that these interactions should 
 play an important role in most systems \citep{Walder14, Cechura15}. 
 
\section{OBSERVATIONS AND DATA REDUCTION} 
\label{observations} 

\subsection{Instrument Setup and Observations} 

\begin{figure}[tb]
\centering
\includegraphics[width=\columnwidth, clip]{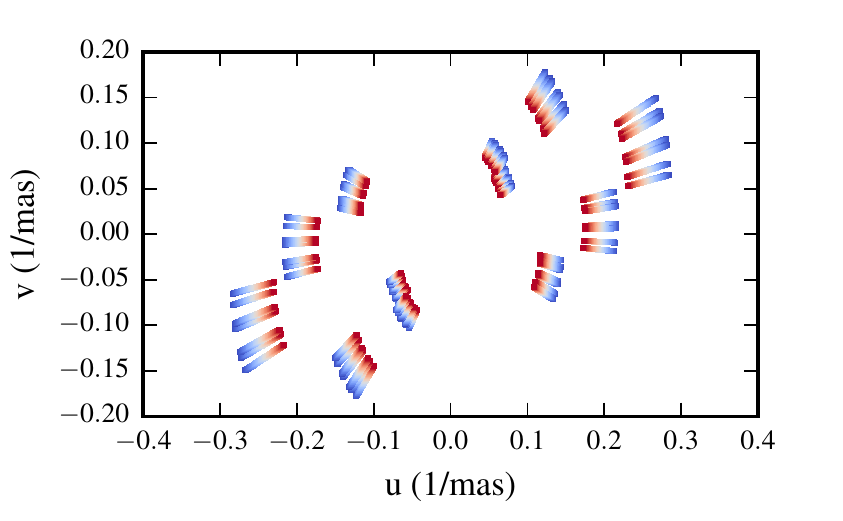} \\ 
\caption{The uv-coverage of our GRAVITY BP Cru observations. The colors represent the different wavelength channels along the K band, from blue ($1.99 \mu$m) to red ($2.45 \mu$m). }
\label{fig: uv coverage}
\end{figure}

We have observed BP Cru with VLTI/GRAVITY on the night of 2016-05-18 with the UTs. 
The observations were carried out in high resolution ($R=4,000$) and in combined (i.e. no split polarization) mode. 
Table \ref{table: observations} summarizes the observations. Figure~\ref{fig: uv coverage} shows the corresponding uv coverage. 

\begin{table}[ht]
\centering
\caption{\label{table: observations} Summary of Observations}
\begin{tabular}{c|c|c|c|c}
\hline \hline
\shortstack{Date\\Time(UTC)} & Mode & \shortstack{Integration\\Time/file} & \shortstack{Total \\Integration\\Time} & \shortstack{Seeing\\(")} \\ \hline
\shortstack{2016-05-18\\00:56-02:14} & \shortstack{HR\\COMBINED} & \shortstack{DIT=30s\\NDIT=10} & $35$min & 0.4-0.6 \\ \hline
\end{tabular}
\tablenotetext{}{}
\end{table}

\begin{figure}[tb]
\centering
\includegraphics[width=\columnwidth, clip]{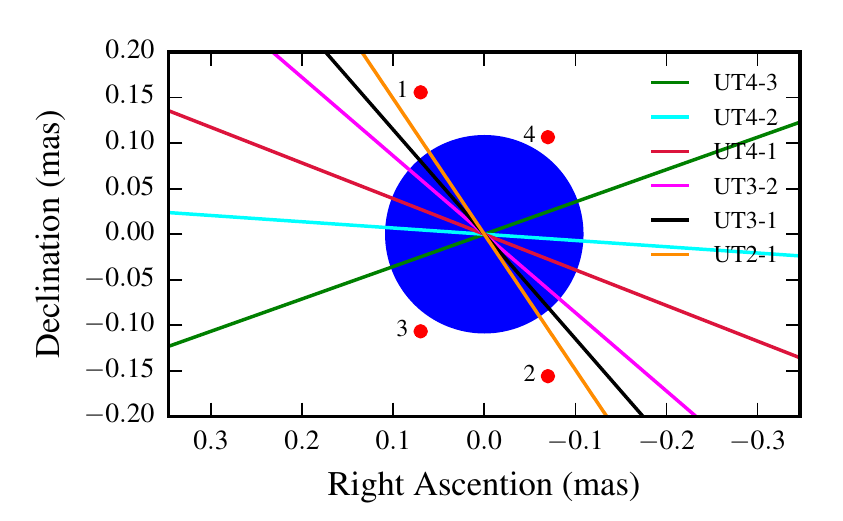} \\ 
\caption{Baseline directions on the sky plane. Also shown are the donor star (photospheric radius $\sim 70 R_{\odot}$) 
and the predicted four possible positions of the pulsar (red) on the sky plane at the time of observation. For details 
see Appendix \ref{app:A}.}
\label{fig: orbit positions}
\end{figure}

The baseline directions on the sky plane are shown in Figure \ref{fig: orbit positions}, together with the predicted binary image at the time
of observation. Because there is no astrometric information on the binary system, the exact position of the pulsar on the sky plane is 
not known. However, we show that we can narrow down its position to the four possibilities shown (see Appendix \ref{app:A}). 

\subsection{Data Reduction}

The data were reduced with the standard GRAVITY pipeline \citep[version 0.9.6,][]{Lapeyrere14}. Most default values were used. In particular, 
$|V|^2$ were debiased and both $|V|^2$ and $|V|$ were corrected for loss of coherence estimated from the FT phase 
deviations using the so called VFACTOR.  

The interferometric calibrators used are listed in Table \ref{table:calibrators}. These stars 
were also used as telluric calibrators for the spectrum. As cool supergiants, they are expected to contain very weak 
absorption lines of hydrogen. In particular, by dividing by an approximate telluric spectrum
\footnote{taken from ESO Spectroscopic Standards \href{http://www.eso.org/sci/facilities/paranal/decommissioned/isaac/tools/spectroscopic_standards.html}{webpage}.},
we checked that there was no remaining Br$\gamma$ line to be removed within the noise level of the spectrum. Unfortunately, the calibrator stars contain CO absorption bands in the red part of the spectrum, which is also affected by telluric lines. Therefore, we do not consider wavelengths $\gtrsim 2.20 \mu$m. This region of the spectrum should not contain any prominent lines for blue hypergiants, and no interferometric signatures are seen in this region.  

\begin{table}[ht]
\centering
\caption{\label{table:calibrators} Interferometric Calibrators}
\begin{tabular}{c|c|c|c|c}
\hline \hline
Name & \shortstack{Spectral\\Type} & \shortstack{Diameter \\(mas)} & Reference & Date \\ \hline
HD 97550 & G8II/III & $0.828 \pm 0.008$ & (a) & 2016-05-18 \\ \hline
HD 110532 & G8Ib/II & $0.804 \pm 0.008$ & (a) & 2016-05-18 \\ \hline
\end{tabular}
\tablenotetext{}{(a) \cite{Merand05}}
\end{table}

The pipeline reports a wavelength calibration with absolute accuracy of $\sim 1$ spectral resolution element ($0.5 \text{ nm}$). 
Since we can achieve statistical errors that are smaller than that when fitting strong emission lines, we cross-correlated 
(IRAF, \textsc{xcsao} package) the uncorrected spectra with the model telluric spectrum in order to reduce the systematic uncertainty 
in the wavelength calibration. We found a global shift $\sim -60 \pm 5 \text{ km/s}$ ($\sim 1$ pixel) consistent for both calibrators and science spectra, and applied the correction.

\section{SPECTROSCOPIC ANALYSIS}
\label{spectrum}

\subsection{Results} 

\begin{figure}[tb]
\centering
\includegraphics[width=\columnwidth, clip]{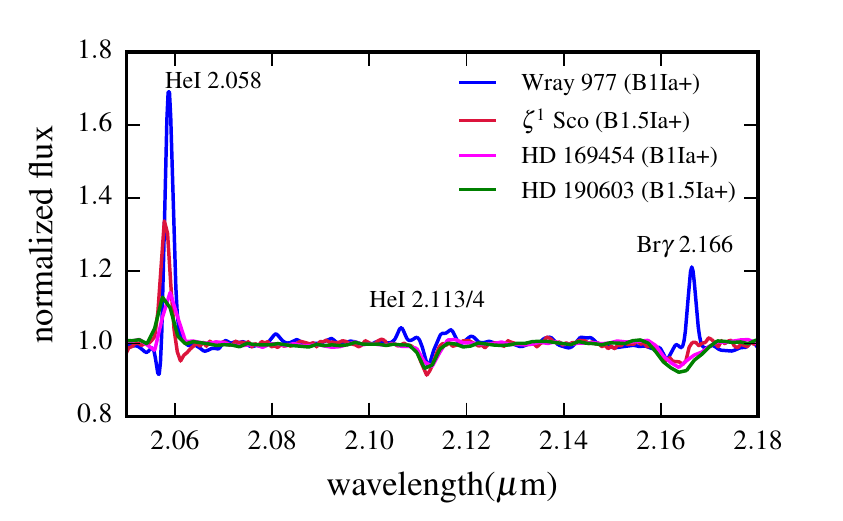} \\ 
\caption{Comparison of Wray 977's GRAVITY spectrum with isolated stars of similar spectral type \citep{Hanson96}. The GRAVITY spectrum has been degraded to the resolution of the $\zeta^1$ Sco spectrum ($R\sim1,500$). The other two spectra have slightly lower resolution, $R \sim 800$. Note the more prominent HeI $2.059 \mu$m emission and the Br$\gamma$ line in emission for Wray 977. The stars have different wind properties, with Wray 977 having the densest wind.}
\label{fig:spectra comparison}
\end{figure}

Currently the most valid spectral classification of Wray 977 is an early blue hypergiant, B1Ia+, based on high-resolution optical spectra \citep{Kaper06}. Figure \ref{fig:spectra comparison} shows part of the K band spectrum obtained with GRAVITY for the UT observations, and comparison spectra of $\zeta^1$ Sco, HD 169454 and HD190603, isolated stars of similar spectral type \citep{Hanson96}. The most striking differences of Wray 977 are its stronger emission in HeI $2.059 \mu$m and Br$\gamma$ in emission rather than absorption. To our knowledge, this is the first published K band spectrum of BP Cru. 

Table \ref{table:spectral lines} shows the identified lines and their measured radial velocities from Gaussian fits (all wavelengths referred are in vacuum). 
The errors shown are purely statistical. In practice, the error is dominated by 
systematic effects caused by the limited spectral resolution and wavelength calibration. 
The velocities were converted to the heliocentric frame. 

\begin{table}[ht]
\centering
\caption{\label{table:spectral lines} Spectral Lines Identified}
\begin{tabular}{c|c}
\hline
\hline
\shortstack{Line\\(Rest Wavelength in Vacuum)} & \shortstack{Measured Velocity\\ (km/s)} \\
\hline
HeI $2.0597 \mu$m & $+29.1 \pm 2$ km/s  \\
HeI $2.1126 \mu$m & $-42.6 \pm 8$ km/s  \\ 
HeI $2.1138 \mu$m & $+1.4 \pm 14$ km/s  \\
Br$\gamma$ $2.1662\mu$m & $+55.4 \pm 4$ km/s \\ 
\hline
\end{tabular}
\end{table}

The double HeI $2.113 \mu$m, $2.114 \mu$m absorption lines are presumably photospheric, and should therefore 
trace the systemic velocity of the system as well as the radial velocity of the supergiant \citep[which is very small, $|v| < 10 \text{ km/s}$,]
[]{Kaper06}. We obtain slightly inconsistent results for the two lines, but this can easily be caused by the limited spectral 
resolution which causes them to be partially blended. A robust result is that the wind emission lines are systematically 
redshifted with respect to the photospheric lines. 

\subsection{Discussion} 

The HeI $2.059 \mu$m line has an unsaturated P-Cygni profile, which suggests an optically thin wind. This line is highly sensitive 
to temperature and wind properties and becomes very active in OB supergiants, acting as a tracer of extended atmospheres \citep{Hanson96}. 
Wray 977 has an estimated mass-loss rate $\sim 5-10 \times$ higher than the comparison stars shown, which is consistent with the stronger
emission. 

The Br$\gamma$ in emission in Wray 977 is a clear deviation from the isolated comparison stars. One explanation could be that its denser wind
drives the line into emission. Unfortunately, these are the only currently known galactic early-B hypergiants of subtype earlier than 2 \citep{Clark12}, so 
this hypothesis cannot be tested observationally. Using detailed stellar atmosphere codes to test this hypothesis is beyond the scope of this paper. Preliminary results (F. Martins, private communication) and previous work \citep{Clark03} suggest that this could indeed be the case. 

Another possibility is that the Br$\gamma$ emission could be caused by denser accretion structures present in the system. As a recombination line, 
Br$\gamma$ emission is usually very sensitive to density \citep{Kudritzki00}. There are many reports 
in the literature of Br$\gamma$ emission lines in X-ray binary systems originating from the accretion disk and its wind. \cite{Shahbaz09} reports 
on a double-peaked Br$\gamma$ emission line for the LMXB V616 Mon, in which the donor star is a K-type dwarf that should not show such an emission line. \cite{Bandyopadhyay99} reports on Br$\gamma$ lines with P-Cygni shape from the LMXB systems Sco X-1 and GX13+1. In the latter, the donor star is 
a K-type giant that is not expected to have emission in Br$\gamma$, whereas in the former the wind terminal velocity is too high to be associated 
with the O-type donor star wind. In both cases, an accretion disk wind is evoked to explain the emission. \cite{Perez09} report on a spectroscopic 
campaign to decompose the Br$\gamma$ emission line of the HMXB and microquasar SS433, and are able to find several emission components, 
including a double-peaked accretion disk component. Also in this case, the A-type donor star supergiant is not expected to show such emission line. 
In several of these cases, HeI lines in the K band, most notably HeI $2.059 \mu$m, are also in emission. 

In HMXBs such as BP Cru, where a stable accretion disk is not expected, associating Br$\gamma$ or HeI line emission with an accretion structure 
is less obvious. However, this possibility should not be excluded in the case of BP Cru, since a gas stream of enhanced density that could be 
dominating the mass-loss rate is expected to be present. The redshifted wind emission relative to the photospheric lines could be explained by 
such a structure or, more generally, by asymmetries in the wind caused by X-ray heating or gravitational disruption by the pulsar.

\section{INTERFEROMETRIC RESULTS}
\label{results}

Here we focus on the main results from the interferometric data. We divide this section in two parts: 
continuum visibilities and spectral differential visibilities. For the purposes of data analysis, the seven 
files were averaged, with the corresponding $(u,v)$ 
coordinates averaged linearly, as appropriate given the short time interval ($\sim$1h20min) spanned by the files. 
Such interval is also negligible compared to the orbital period and X-ray variability timescale. 

\subsection{Continuum Size and Asymmetry}
\label{continuum size}

Here we estimate an upper limit on the continuum size from the continuum visibility amplitudes. 
The most reliable visibility amplitude estimator is the squared visibility modulus of the fringe 
tracker (FT), since it measures the fringe visibility within the coherence time of the atmosphere. 

We note that the continuum closure phases are zero to within the noise limit ($\text{RMS}<1\degr$) on all baselines. The closure
phase is much more robust to systematic errors than the visibility amplitudes, and therefore there is strong indication for a 
symmetrical continuum emission. Since, in addition, the source is very close to unresolved, there is no big difference between using a 
disk, Gaussian or any similar model for the continuum $|V|^2$. We choose a uniform disk model with the angular diameter as the only parameter. 

\begin{figure}[tb]
\centering
\includegraphics[width=\columnwidth, clip]{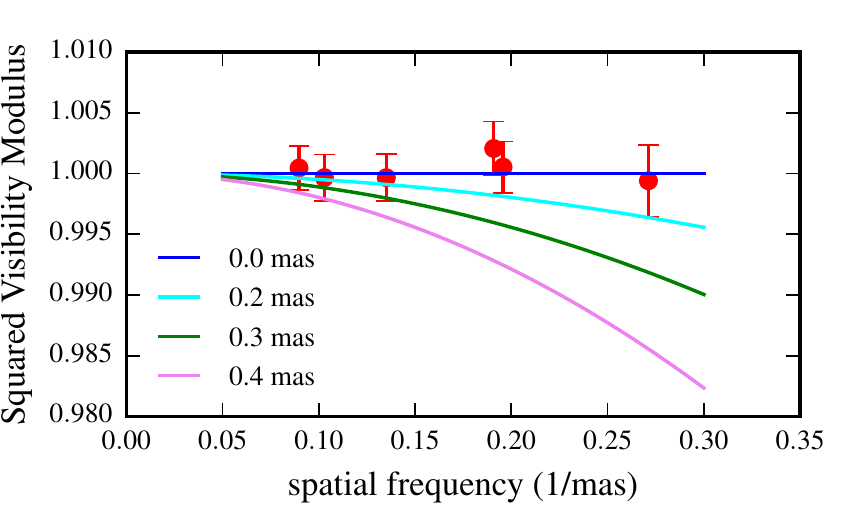} 
\caption{Continuum visibility amplitudes (spectrum average) measured by the fringe tracker. Disk models with varying angular diameters are shown for comparison.}
\label{fig:continuum}
\end{figure}

Figure \ref{fig:continuum} shows the squared visibility modulus measured by the FT, averaged over the five spectral channels for each 
baseline. The error bars include the measurement errors from the science object, as well as from the calibrator object and the calibrator 
diameter's systematic uncertainty $\sim 1\%$. Disk models with the indicated angular diameters are also plotted for comparison. 

The data is most consistent with an unresolved continuum of size $\theta_d \lesssim 0.2 \text{ mas}$. Because the continuum size is in the very challenging limit that is well below the interferometer canonical resolution $\theta \ll \frac{\lambda}{|\vect{B}|} \sim 3$ mas, 
the measurements are very sensitive to systematic errors between baselines. We therefore restrain from a formal fit, and restrict to providing a very conservative upper
limit to the continuum size $\theta_d \lesssim 0.4 \text{ mas}$. Structures larger than this are clearly inconsistent with the data, as shown in Figure \ref{fig:continuum}. 
 
\subsection{Differential Visibilities and Phases}

For treating the differential visibility signatures, we averaged the seven files after normalizing the 
visibility amplitudes to an unresolved continuum. The visibility phases are 
output from the pipeline already mean and slope subtracted i.e. as differential quantities. 

\begin{figure*}[tb]
\centering
\includegraphics[width=2.2\columnwidth, clip]{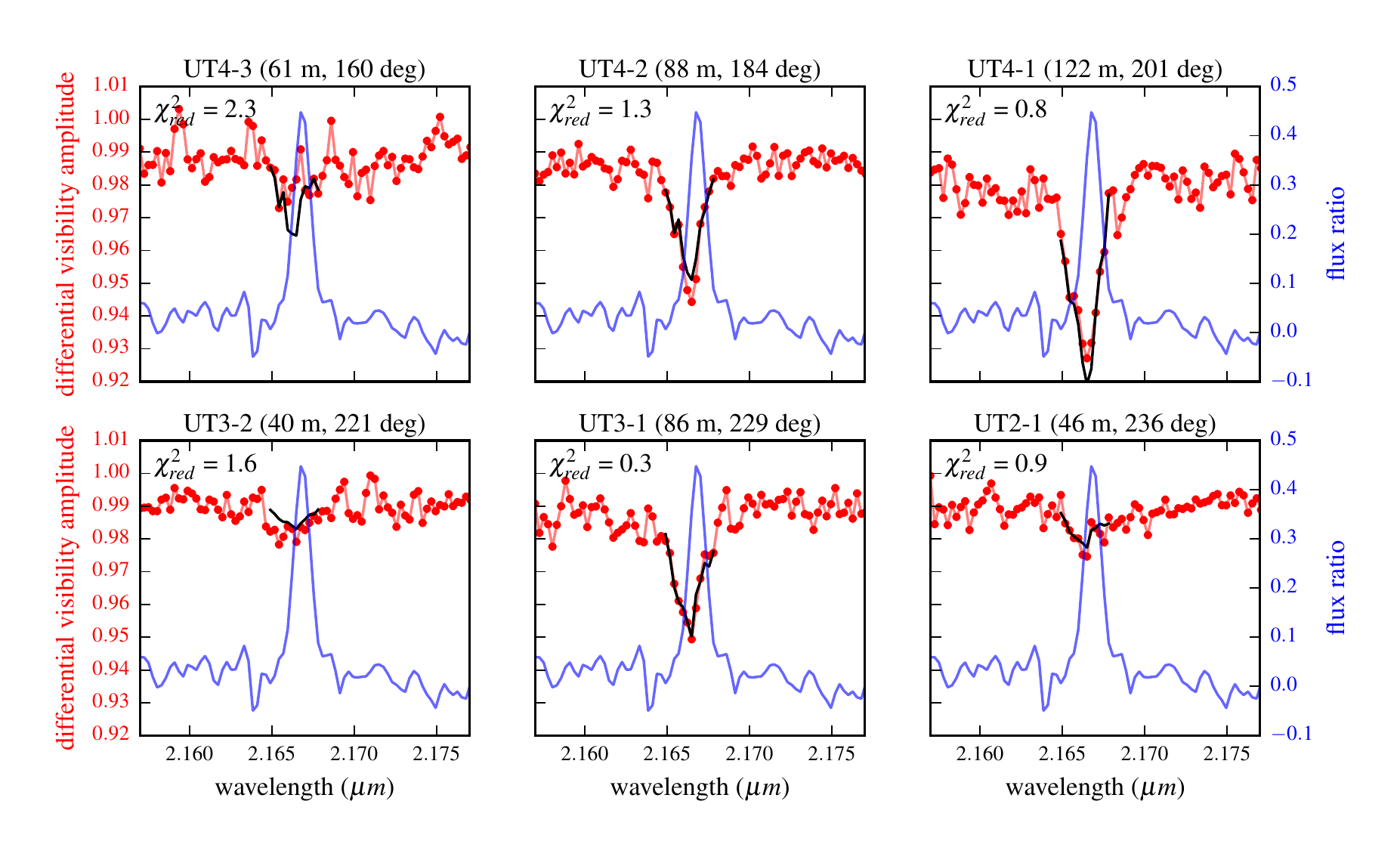} \\ 
\caption{Differential visibility amplitudes at Br$\gamma$ line (red) and the normalized photospheric-corrected flux ratio (blue). For each baseline, the projected baseline length and the position angle are also shown. In black, we show model-independent fits to the visibility amplitudes (see text for details).}
\label{fig:diff visamp HI}
\end{figure*}

\begin{figure*}[tb]
\centering
\includegraphics[width=2.2\columnwidth, clip]{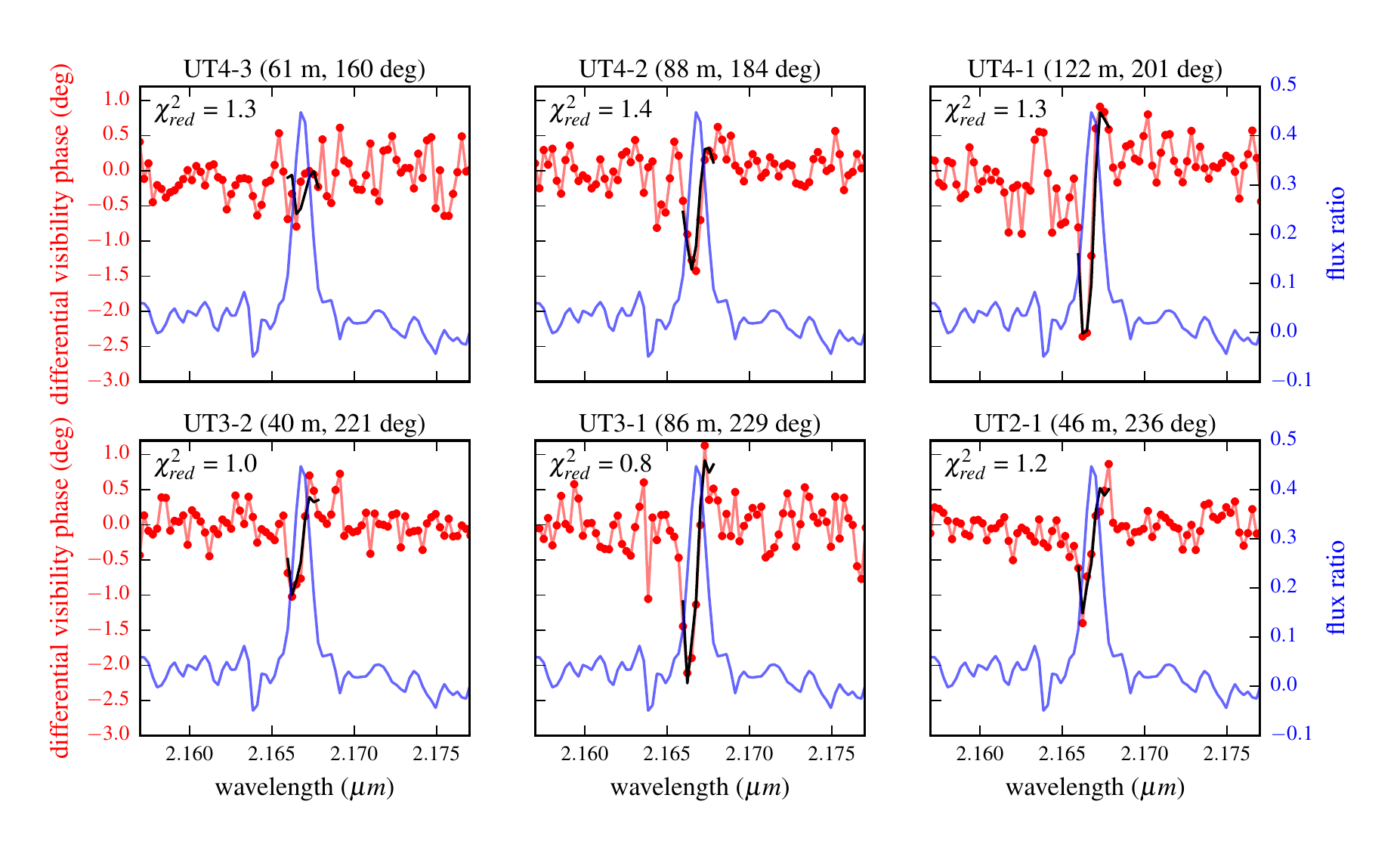} \\ 
\caption{Differential visibility phases at Br$\gamma$ line (red) and normalized photospheric-corrected flux ratio (blue). For each baseline, the projected baseline length and the position angle are also shown. In black, we show model-independent fits to the visibility phases (see text for details).}
\label{fig:diff visphi HI}
\end{figure*}

Figure \ref{fig:diff visamp HI} shows the differential visibility amplitudes across the Br$\gamma$ line for the six baselines at hand. 
The photospheric-corrected flux ratio 
(see Appendix \ref{app:B}) between the continuum and the line emission is also shown for comparison. In general, 
the visibility amplitudes show, for some baselines, a decrease at the lines relative to the continuum, which 
is indicative of extended or multi-component emission. However, the peak of the $|V|$ drop does not happen at 
the center of the line, but rather it is displaced to the blue side. Figure \ref{fig:diff visphi HI} shows the differential visibility phases. They show larger, 
negative values on the blue side of the line and, for some baselines, smaller, positive values on the red side of the line. Such "S-shaped"
differential visibility signatures across a line are typical interferometric tracers of rotation \citep[e.g., they are often observed 
in Be stars, in which they are attributed to extended equatorial disks, but in these systems the blue and red phase signatures are roughly symmetric,][]{Meilland12}. The black lines in the plots are 
model-independent fits to the data and will be discussed in the following section. 

\begin{figure}[tb]
\centering
\includegraphics[width=\columnwidth, clip]{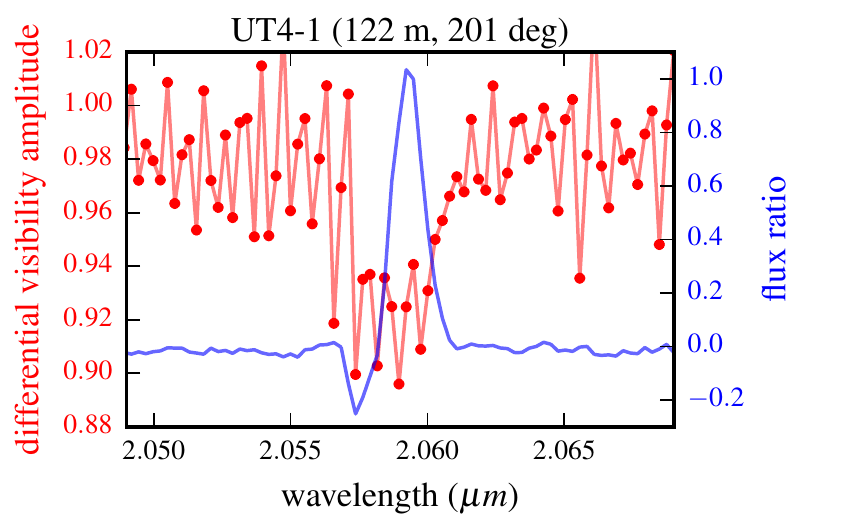} \\ 
\includegraphics[width=\columnwidth, clip]{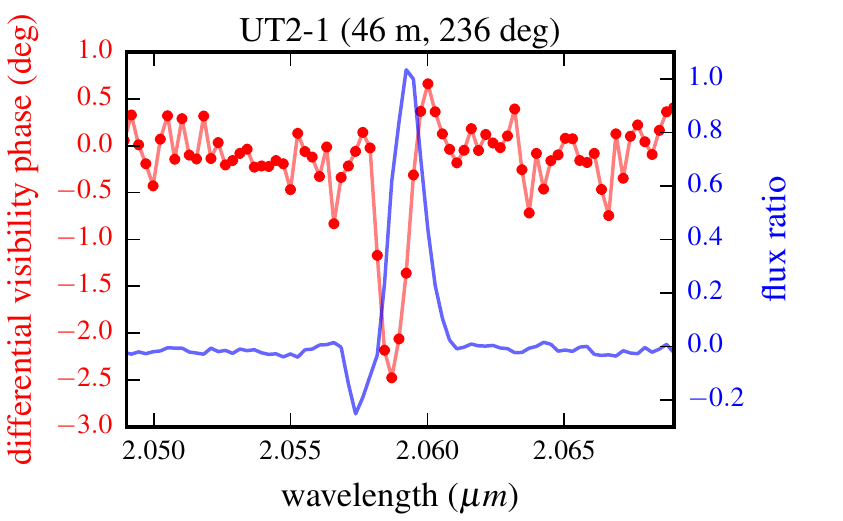} \\ 
\caption{Differential visibility amplitudes and phases across the HeI $2.059 \mu$m line for some representative baselines. The features 
agree with those seen in Br$\gamma$, but are, in general, noisier due to instrumental and atmospheric effects.}
\label{fig:HeI}
\end{figure}

Similar interferometric features in both differential visibility amplitudes and phases are also found across the HeI $2.059 \mu$m 
line. However, this region of the spectrum suffers from a particularly high level of noise due to the GRAVITY metrology laser and 
the large telluric absorption. For instance, the RMS in the visibility amplitude, estimated from the scatter in the continuum region 
around the lines, is $0.4 \%$ and $1.2 \%$ for Br$\gamma$ and HeI, respectively. Similarly, the corresponding values for 
differential visibility phases are $0.2 \degr$ and $0.6 \degr$. That, in addition to the more complicated (P-Cygni) shape of the 
line, led us to focus our analysis on the Br$\gamma$ line. We show in Figure\ref{fig:HeI} the visibility signatures across the HeI $2.059 \mu$m 
line for some representative baselines. 

Several factors point to the credibility of such features. The wavelength alignment between the extracted spectrum for each telescope agrees to $< \frac{1}{2}$ of a resolution element. Similar features are not found at other lines in the spectrum, either related to the science object (e.g. HeI 2.113/4$\mu$m) or telluric. Moreover, they show up with different strengths for different baselines (as expected for any reasonable interferometric model) and are consistent between the two emission lines. Finally, for the differential visibility amplitudes, the features are strongest in three baselines which encompass all of the four telescopes, whereas for the differential visibility phases a signature is detectable in five of the six baselines.  

\subsection{Closure Phases} 

Closure phases are sums of visibility phases formed in a closed triangle of baselines which are independent of telescope errors. For this reason they are robust probes of asymmetry. As mentioned above, the closure phases across the continuum are zero to within the noise on all four baseline triangles (only three are independent). In theory, \textit{differential} closure phases are not independent measurements 
from what has already been presented since they are derived from linear combinations of differential phases. 

\begin{figure*}[tb]
\centering
\includegraphics[width=2.2\columnwidth, clip]{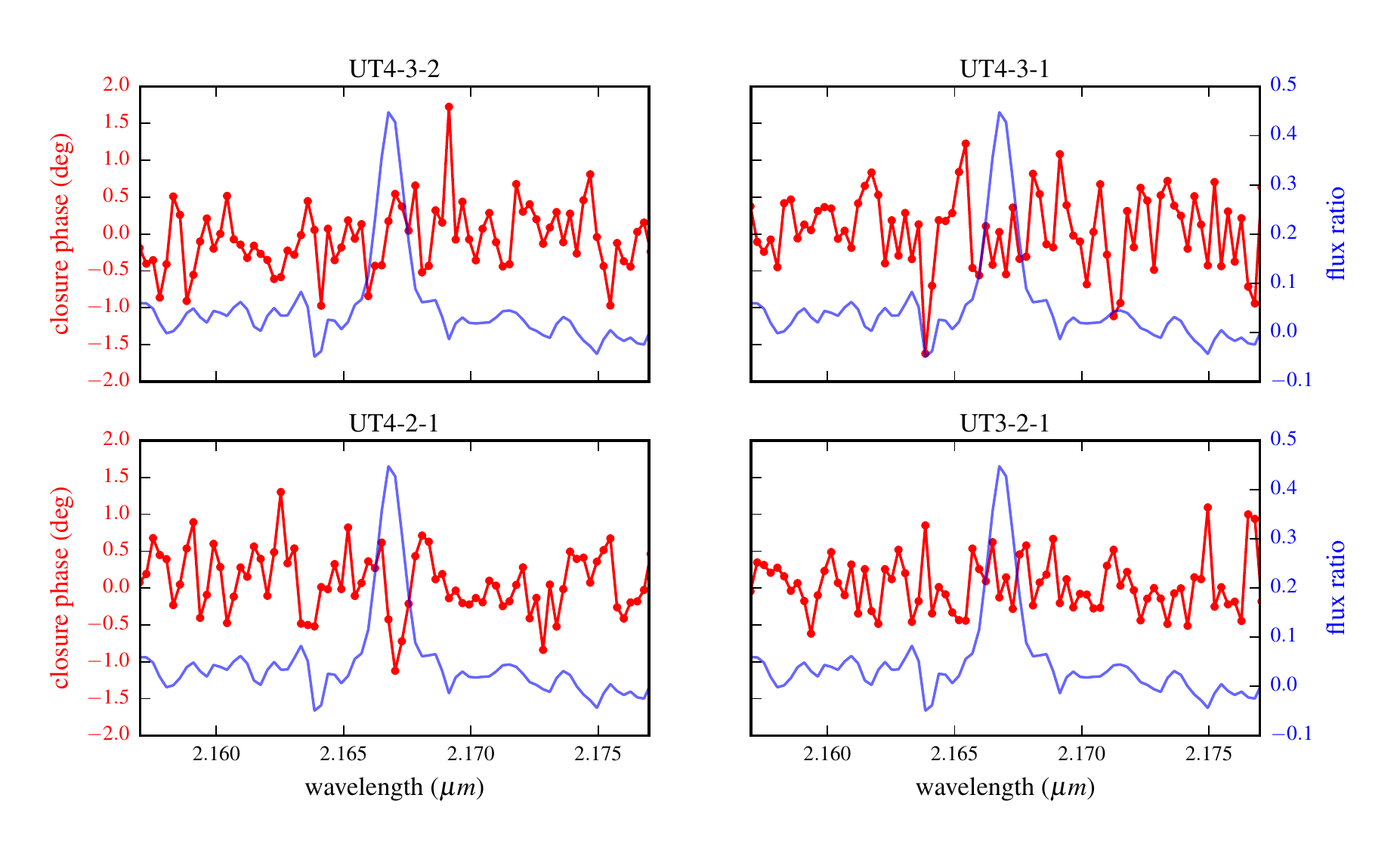} \\ 
\caption{Differential closure phases across Br$\gamma$ line (red) and normalized photospheric-corrected flux ratio (blue). In contrast to the 
differential visibility phases, there is no clearly distinguishable feature within the noise.}
\label{fig:t3phi HI}
\end{figure*}

Figure \ref{fig:t3phi HI} shows the differential closure phases 
across the Br$\gamma$ line, which also vanish to within the noise level. Even though the differential closure phases are naturally noisier than 
the individual baseline differential visibility phases by $\approx \sqrt{3}$ (RMS$=0.4 \degr$), the fact that they vanish might be puzzling at first since the differential 
visibility phases are non-zero and therefore indicate the presence of asymmetry. This will be 
clarified in Appendix \ref{app:A}.

\section{Discussion} 
\label{discussion} 

\subsection{Continuum} 

The photospheric radius $R(\tau_{Ross}=2/3) = 70 R_{\odot}$ and the distance $3 \text{ kpc}$ to Wray 977 \citep{Kaper06} imply 
a photosphere angular diameter $\theta \approx 0.2 \text{ mas}$. Our continuum size measurements are therefore consistent with a size $\lesssim 2\times$ the photosphere diameter, using 
our conservative upper limit referenced above. For hot stars with strong winds, the observed continuum emission in the infrared is a combination of blackbody thermal emission around the photosphere region as well as bound-free and free-free emission in the optically thin wind. \cite{Kaper06} compares the SED of Wray 977 with a Kurucz model with the same temperature and finds a strong infrared excess, associated with emission from the wind. However, at the maximum wavelength probed by GRAVITY,  $\sim 2.5 \mu$m, the wind contribution is still relatively small, $\sim 20\%$ of the flux. Therefore, it is expected that the continuum in the K band is still dominated by the photosphere rather than the wind. This is consistent with the interferometric results presented here. Furthermore, the lack of a resolved structure in the near-infrared continuum also argues against the presence of a circumstellar disk, which is often seen in Be stars as extended continuum emission in the K band with FWHM $\gtrsim 2 D_*$ \citep{Meilland12}. 
 
\subsection{Differential Visibilities} 

The main advantage of using spectral differential visibility measurements is that they are much less susceptible to systematic errors that can affect the absolute visibility quantities. The errors in fringe contrast and phase are, in general, monotonic functions of the phase difference caused by spurious OPDs between baselines, $\Delta \phi = \frac{2\pi}{\lambda}OPD$. The error in the differential quantities will then have the form $f (d\Delta \Phi) \approx f(-2 \pi \frac{OPD}{\lambda} \frac{d\lambda}{\lambda})$, which is greatly reduced with respect to the non-differential error when $\frac{d\lambda}{\lambda} \ll 1$, which is the case, for example, when using the wavelength of a narrow line compared to the continuum around it. On top of that, the differential quantities are not affected by wavelength-independent errors and are robust to low-order spurious effects along the spectrum given the narrowness of the spectral lines. 

\subsubsection{Model-independent Analysis in the Marginally Resolved Limit} 

The downside of spectral differential quantities is that, when imaging is not possible, their ultimate interpretation relies on knowing the spectral decomposition of the line, in case there is more than one emission component. Given the likely complex nature of the 
source in question and the many possible components in the system (hypergiant photosphere, wind, pulsar, gas stream, accretion disk etc), it 
would be useful to derive model-independent properties about the image that any model would have to reproduce. In general, this is not 
possible without image reconstruction, which requires a much more dense u-v sampling than we have available here. 

However, when the interferometric signatures are small, such as is the
case here, spectral differential quantities nicely fit into the
special framework of the marginally resolved limit in
interferometry. \cite{Lachaume03} lays out the formalism of this limit
focusing on absolute visibilities and closure phases, showing that the
visibility signals can be related to the moments of the flux
distribution in a model-independent way. We present a similar analysis in Appendix
\ref{app:C}, focusing on spectral differential signatures. In summary,
the validity of this limit for this data set is confirmed by large
visibility amplitudes $|V| > 90\%$, small ($<3 \degr$) differential
visibility phases and vanishing closure phases (or closure phases that
are much smaller than the individual visibility phases). Our data
satisfy all 3 conditions.

In this context, as shown in appendix \ref{app:C}, the differential
visibility phases are probes of the difference of the centroid position of the
image at the continuum and the image at a spectral channel within the
line (which includes emission from both the continuum and the
line). Therefore, there are only two parameters and they can be fit
(per spectral channel) to the six baselines. The best fit model (and
corresponding $\chi^2_{red}$) are shown in Figure \ref{fig:diff visphi
  HI} (black line). For this and all subsequent model fits, we use a
Markov Chain Monte Carlo technique as implemented in the publicly
available \textsc{emcee} code \citep{Foreman13} using uniform priors. We only
fit the spectral channels in which there is emission line flux above
the continuum noise level. 

\begin{figure}[tb]
\centering
\includegraphics[width=\columnwidth, clip]{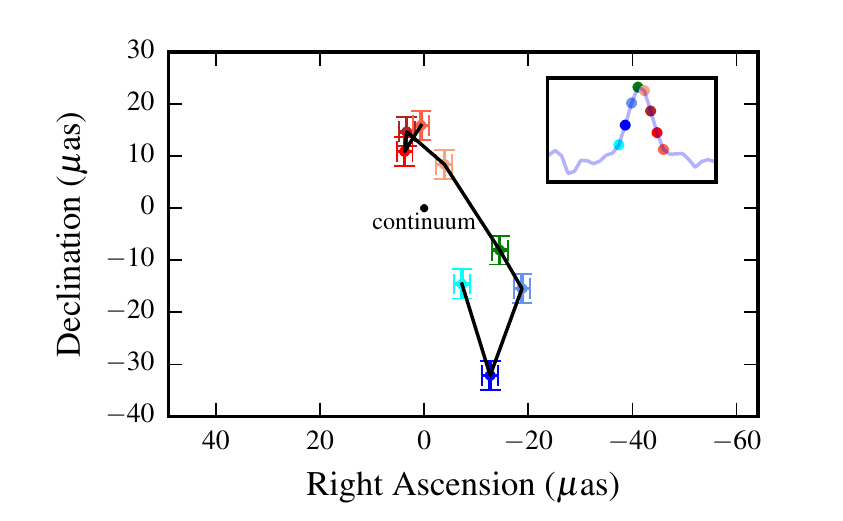} \\ 
\includegraphics[width=\columnwidth, clip]{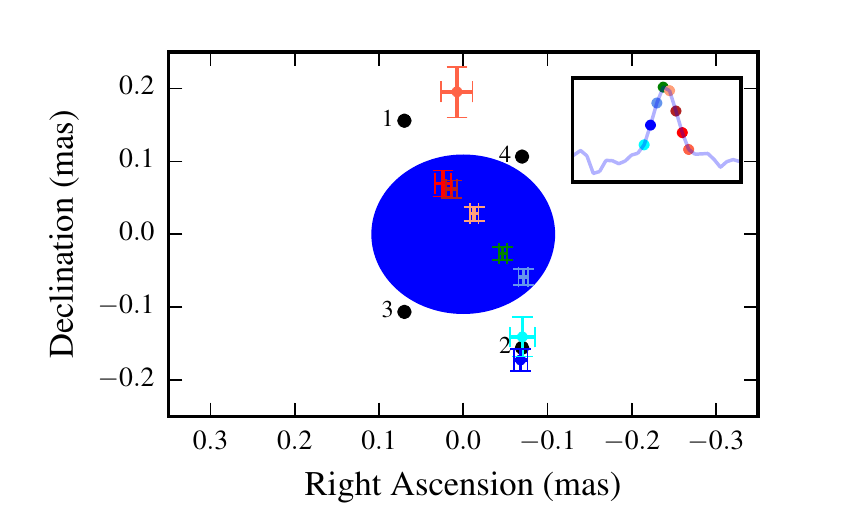} \\ 
\caption{\textbf{Top.} Model-independent centroid positions for each wavelength across the Br$\gamma$ line (continuum is at (0,0)). The image on the blue side of the line has a larger centroid shift 
as compared to the image on the red side. \textbf{Bottom.} Same as above, but using the flux ratio to derive the barycenter of the line emission. The hypergiant and the predicted four possible pulsar positions are also shown.}
\label{fig:visphi HI sky}
\end{figure}

The consistency between the six baselines is further confirmation that the marginally resolved limit is valid. The resulting centroids on the sky plane for each wavelength across the differential signature are shown in Figure \ref{fig:visphi HI sky}. The errorbars shown correspond to the $16\%$ and $84\%$ marginalized quantiles. 
The red part of the line must have a smaller ($\sim 10 \mu$as) centroid shift with respect to the continuum image than the blue part of the line ($\sim 30 \mu$as). This statement is model-independent. Because the image at the line contains both a line as well as a continuum contribution, we can estimate the barycenter of the line emission with respect to the continuum (at (0,0)) by scaling the model-independent centroids by $\frac{1+f}{f}$, where $f$ is the flux ratio between continuum and line emission (see Appendix \ref{app:B}). This, however, must be interpreted carefully since the line emission could have more than one component. The result is also shown in Figure \ref{fig:visphi HI sky}. The resulting centroid positions suggest line emission offset from the continuum by less than the size of the binary orbit, with a spatial gradient across wavelengths and the bluest channels consistent with one of the possible positions of the pulsar on the sky plane.  

\begin{figure}[tb]
\centering
\includegraphics[width=\columnwidth, clip]{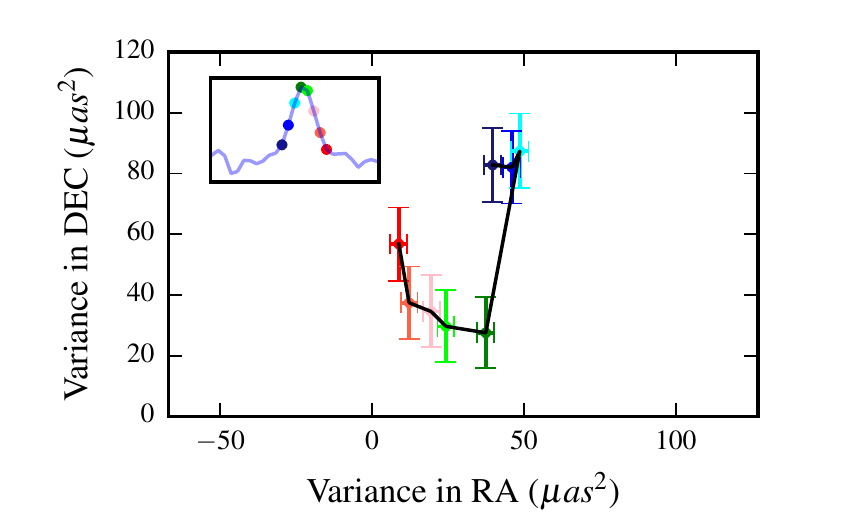} \\ 
\caption{Model-independent variances of the image as a function of wavelength across the Br$\gamma$ line. The blue part of the line has higher values, which suggests that the emission must be coming from larger scales.}
\label{fig:visamp HI var}
\end{figure}

As shown in Appendix \ref{app:C}, in the marginally resolved limit the differential visibility amplitudes carry model-independent information about the difference in the second-order moments (variances and covariance) about the centroids of the image at the continuum and the image at the spectral channel within the line. Therefore, there are three parameters to be fit at each wavelength for six baselines. Analogously to the differential phase case, the fit results are shown in Figure \ref{fig:diff visamp HI} (black line). Again, a consistency between baselines confirms the validity of the marginally resolved limit. The resulting variance difference is both RA and DEC as a function of wavelength is shown in 
Figure \ref{fig:visamp HI var}. Clearly, a higher variance is required on the blue side of the line, implying that this part of the emission must come from larger scales. Also, the fact that the variances are not symmetrical in RA and DEC suggests an asymmetric emission structure. 

\begin{figure}[tb]
\centering
\includegraphics[width=\columnwidth, clip]{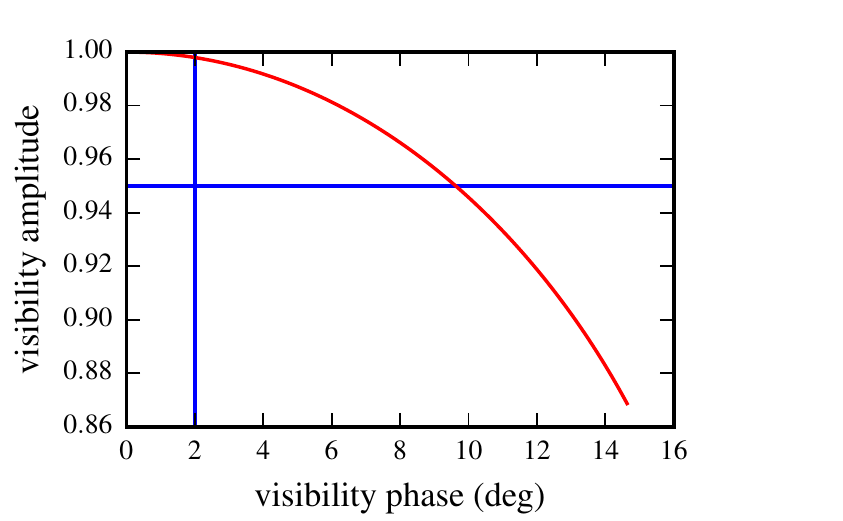} \\ 
\caption{Visibility amplitude vs phase as the separation is changed for a 1D binary model with flux ratio $f=0.3$ and u coordinate $0.2 \text{ mas}^{-1}$. The measured visibility amplitudes $\sim 95\%$ and phases $\sim 1-2 \degr$ are not compatible with this simple model.}
\label{fig:signature comparison}
\end{figure}

The differential amplitude signatures are larger than expected from the differential phases. For example, for a binary model with compact components and flux ratio given by the spectrum, the binary separation as implied by the differential phases is one order of magnitude smaller than the one that would be necessary to produce the differential visibility amplitudes. This is illustrated in Figure \ref{fig:signature comparison}, where we plot the visibility amplitude vs phase for a 1D binary model as the binary separation is changed. We choose a flux ratio $f = 0.3$ and a u-coordinate $0.2 \text{ mas}^{-1}$, which are appropriate to our data. We can clearly see that visibility amplitudes $\sim 95\%$ are not compatible with visibility phases $\sim 1-2 \degr$. This statement is robust and not dependent on the chosen $f$ and $u$. 

\subsubsection{Simple Geometric Models}

The model-independent analysis in the context of the marginally resolved limit 
presented above allows to derive properties that any interferometric model has to satisfy in order to explain the data. In summary: 

\begin{enumerate}
\item the image centroid must have a spatial gradient across the spectrum, with larger centroid deviations from 
the continuum at the blue side of the line, and in the opposite direction at the red side; 
\item the image variance must also show such a gradient, with larger spatial extension also at the blue side of the line; and
\item small centroid displacements must coexist with large scale structure.
\end{enumerate} 

Fitting the data with complex hydrodynamic models which produce Br$\gamma$ emissivity maps is beyond the  scope of this paper. Instead we restrict ourselves to the use of physically motivated, geometric 
models. We note that any interferometric model must deal with flux ratios, which are often 
degenerate with the spatial parameters. Whereas the simplest assumption is to use the spectrum to set the flux 
ratio, this only works if there is only one emission component. Since determining a complex spectral decomposition 
from interferometric data at moderate resolution is not possible, we limit ourselves to the simplest assumptions in the following models. 

\subsubsection*{Model A: Extended and Distorted Wind}

In this model, we assume that the Br$\gamma$ emission is completely dominated by the hypergiant stellar wind. A spherically symmetric wind 
centered on the star would not be able to produce differential visibility phases with respect to the continuum; therefore, we allow the wind, which 
is modeled as a Gaussian, to be displaced from the center. For each wavelength channel across the Br$\gamma$, we therefore model
the complex visibility as
\begin{equation}
V (\vect{u}) = \frac{V_{cont}(\vect{u}) + f e^{-\pi^2 |\vect{u}|^2 \frac{\theta_d^2}{4 \log 2}} e^{-2 \pi i \vect{\sigma_0} \cdot \vect{u}}}{1+f}
\end{equation}
where $V_{cont}(\vect{u})$ is the continuum visibility, $f$ is the photospheric-corrected flux ratio between wind emission 
and continuum set by the spectrum, and the fit parameters are $\theta_d$, the FWHM of the wind, and $\vect{\sigma_0}$, the centroid position
of the wind. 

\begin{figure}[tb]
\centering
\includegraphics[width=\columnwidth, clip]{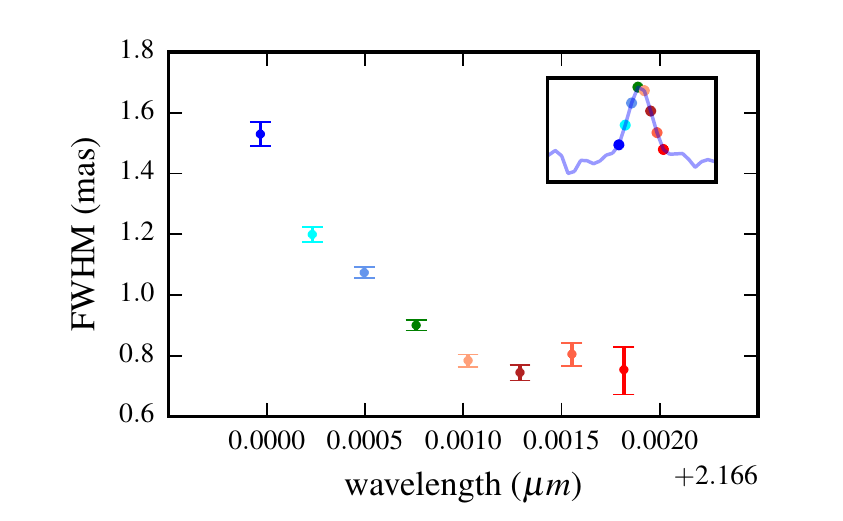} \\ 
\caption{Wind size (FWHM) as a function of wavelength for a model in which the Br$\gamma$ emission is dominated by the wind. Such a model predicts 
that there is still substantial wind emission at $4-7 \times R_*$, and that the blue (approaching) part of the wind is up to $\sim 2 \times$ more 
extended than the red (receding) part.}
\label{fig:wind sizes}
\end{figure}

This model is fit to both visibility amplitudes ($\chi^2_{red} = 2.67$) and differential visibility phases ($\chi^2_{red} = 1.36$). Because the (differential)
closure phases can be derived from the visibility phases, they are not included in the fit; in other words, a good fit with respect to differential visibility phases
should automatically be consistent with differential closure phases. The resulting centroid fits are identical to those shown in Figure \ref{fig:visphi HI sky} (bottom), 
as they should, since we are likewise assuming here that only one (spherically symmetric) structure contributes to the emission. 
The resulting wind sizes, as a function of wavelength, are shown in Figure \ref{fig:wind sizes}. 

The resulting wind FWHM (from $\sim 0.8 \text{ mas}$ on the red part of the wind up to $\sim 1.5 \text{ mas}$ on the blue part) would imply that there is 
substantial emission in Br$\gamma$ up to $\sim 4-7 \times R_*$. On the other hand, the non-Lyman H lines in hot stars are usually recombination lines, which
means that their source function is roughly Planckian and stays approximately constant throughout a wind that is at radiative equilibrium. At the same 
time, their opacity $\kappa \propto \rho^2$ is a very sensitive function of density, and for an accelerating wind with a fast-decaying density profile
($\rho \propto \frac{1}{r^2 v(r)}$), only the innermost ($\sim 1-1.5 R_*$) regions of the wind would have a substantial contribution to the emission 
\citep{Kudritzki00}. A varying temperature profile and the dependence of optical depth with velocity gradient ($\tau \propto \frac{dv}{dr}$) might smooth 
the density decay, but it is unlikely to resolve the discrepancy in the case of Wray 977, where the CAK wind law \citep{Castor75} predicts a density at $4 R_*$ that is already
$\sim \frac{1}{1000}$ of the value at $R_*$. A radiative transfer calculation to determine the emission region of Br$\gamma$ in the wind is beyond the 
scope of this work; nonetheless, preliminary results (F. Martins, private communication) show that a dense wind could indeed bring Br$\gamma$ into emission, 
but the emission region would be sharply peaked between $\sim 1.3-3 R_*$, therefore unable to account for such extended emission. Mid-infrared observations of BP Cru have detected the presence of dust and the possibility that the binary system is enshrouded by a disk-like circumstellar envelope $\sim 2$\,mas \citep{Servillat14}. Even though (i) the optical spectrum shows no evidence for a circumstellar disk (ii) the interferometric signatures are not typical of a symmetric disk and (iii) the near-infrared continuum is unresolved, there could be a connection between the very extended wind emission seen in these data and the reported dusty CS structure in the mid-infrared. 

Another feature of the wind model is that the blue (approaching) side of the wind would have to be $\sim 1.5-2 \times$ more extended than the red (receding) part, where the pulsar is predicted to be at the time of the observation. This could be due to the X-ray illumination of the red part of the wind that hinders the radiative acceleration of the wind by photoionization. 

The centroid shifts of the wind with respect to the continuum, necessary to explain the differential visibility phases, are small with respect to the size of 
the wind, $|\vect{\sigma_0}|/\theta_d \sim 10\%$. Because a Gaussian image 
has no intrinsic phase, the small centroid shifts in the model might 
be indicative of asymmetric wind structure. Such asymmetries could arise from a clumpy wind, or, more generally, from density fluctuations in the wind, which could be caused by the influence of the gravitational or radiation fields of the compact object. Although \cite{Kaper06} found no evidence for wind clumping in 
Wray 977 from optical spectrum modeling, X-ray light curves and column density measurements often show fluctuations potentially attributed to clumps 
in the stellar wind \citep{Leahy08}. 

We also recall that the interferometric data on Vela X-1 \citep{Choquet14}, whose supergiant also possesses a strong wind, did not find any differential visibility signatures at the spectral lines above the noise level. GRAVITY commissioning data on this same target also had the same conclusion, even though the SNR was comparable to the one here (RMS in differential visibility amplitudes and phases in the continuum around the Br$\gamma$ line were $1.2\%$ and $0.7 \degr$, respectively). However, the donor star in Vela X-1 is $\sim 2 \times$ smaller and has a $\sim 5 \times$ smaller mass loss rate than Wray 977, and the spectral lines in K band are in absorption or very weak emission.

\subsection*{Model B: Extended Wind + Gas Stream}

Here we consider the possibility that a gas stream of enhanced density also contributes to the Br$\gamma$ emission. 
The manifestation of a gas stream of enhanced density in the hydrogen emission lines of HMXBs is not completely unfamiliar. \cite{Yan08} 
 for e.g. studied the double-peaked  H$\alpha$ emission lines in Cyg X-1, which can be explained by a P-Cygni shaped
 wind profile that follows the orbit of the supergiant as well as emission from a focused stellar wind that has an approximately anti-phase orbital motion
 to the supergiant. The relevance of the focused wind in Br$\gamma$ could be even higher than in H$\alpha$ given that the former line requires much higher 
 densities to be brought into emission. 

As alluded above, a gas stream is predicted to be present in this system from both optical and especially X-ray data. Because of its
compactness, a gas stream could also be more efficient than a stellar wind in bringing higher density regions to the outer parts of the 
system. The simplest stream model would therefore be a binary model consisting of the continuum region at the center and an extra unresolved component. 
However, it was already shown that a binary model cannot explain the discrepancy between the very small differential visibility phases and the larger 
differential visibility amplitudes. This is confirmed in a formal binary fit to the data, which is completely unsatisfactory in reproducing both visibility amplitudes 
and phases simultaneously. 

\begin{figure}[tb]
\centering
\includegraphics[width=\columnwidth, clip]{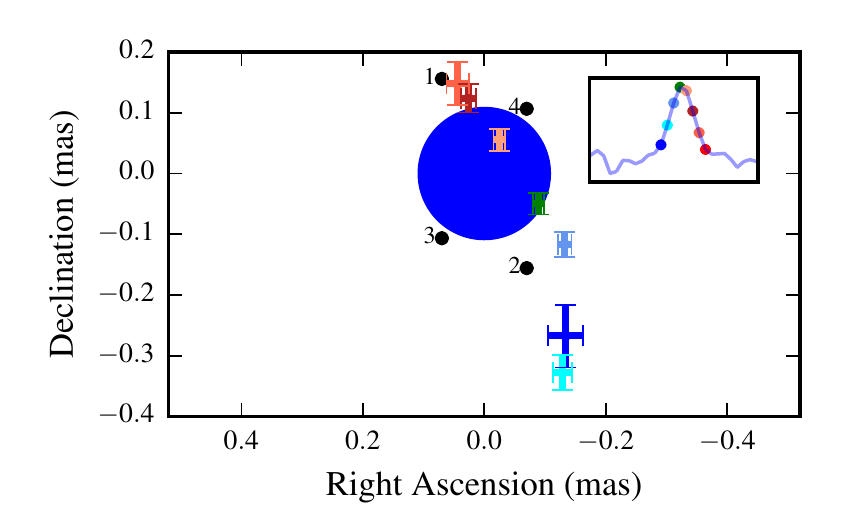} \\ 
\includegraphics[width=\columnwidth, clip]{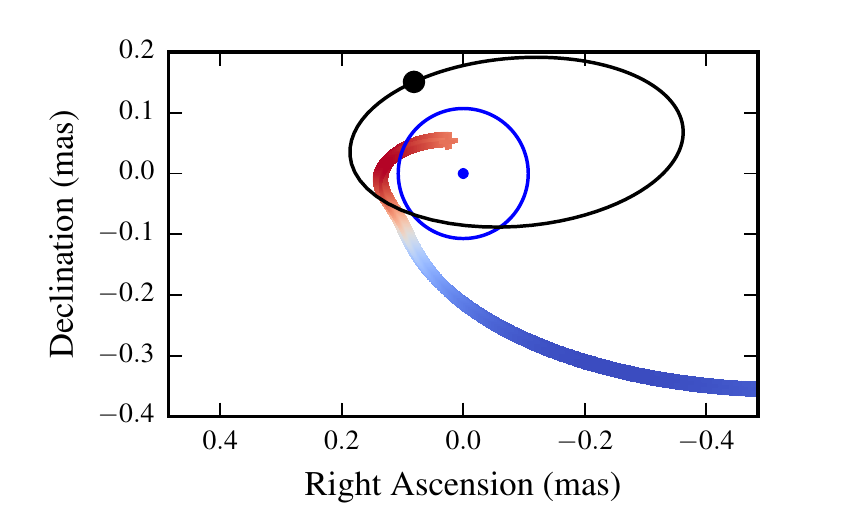} 
\caption{\textbf{Top:} Best fit positions on sky plane for a gas stream in the combined wind+stream model. 
Also shown are the hypergiant and the predicted four possible positions of the pulsar. \textbf{Bottom:} Example of 
a gas stream model \citep{Leahy08} in the sky plane. The colors refer to radial velocities. A gas stream could be 
an explanation for asymmetric differential visibility phases across the wavelength.}
\label{fig:combined sky}
\end{figure}

\begin{figure}[tb]
\centering
\includegraphics[width=\columnwidth, clip]{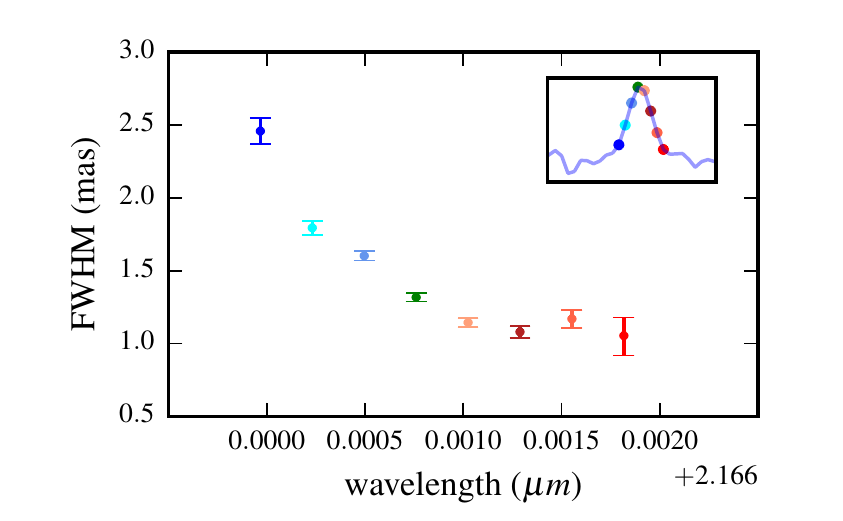} \\ 
\caption{Wind Size (FWHM) as a function of wavelength for a combined wind+stream model. The asymmetry in extension across 
wavelength remains, as in the wind-only model.}
\label{fig:combined size}
\end{figure}

Motivated by this discrepancy, we consider here the possibility that the Br$\gamma$ line has two emission components: a gas stream of enhanced density, with 
size on the order of the orbit scale and which accounts for the asymmetric differential visibility phase signatures, and an extended wind, which is symmetric relative to the 
continuum and accounts for most of the differential visibility amplitude signatures. Because of the lack of higher spectral resolution, it is not possible to perform a spectral decomposition
to fix the flux ratios for each component. Because the flux ratios are highly degenerate with the spatial parameters, we fix them to be equal for the stream and wind components. 
This is motivated by comparing the HeI $2.059 \mu$m line in Figure \ref{fig:spectra comparison} for BP Cru and $\zeta^1$ Sco: they have similar stellar parameters, so if the extra emission 
is due to a stream, it would account for roughly $50\%$ of the line emission. We caution that Br$\gamma$ and HeI $2.059 \mu$m have very different behavior, and the goal of this section 
is not to provide best fit parameters, but rather to assess the possibility of a combined wind+gas stream model. Furthermore, we assume that the Br$\gamma$ emissivity is constant along
the stream, which might not be the case. The complex visibility at each spectral channel is therefore modeled as 
\begin{equation}
V (\vect{u}) = \frac{V_{cont}(\vect{u}) + \frac{f}{2} e^{-\pi^2 |\vect{u}|^2 \frac{\theta_d^2}{4 \log 2}} + \frac{f}{2} e^{-2 \pi i \vect{\sigma_1} \cdot \vect{u}}}{1+f}
\end{equation}
where all parameters are as in Model A and $\vect{\sigma_1}$ is the position of the stream. Figure \ref{fig:combined sky} (top) shows the positions of the stream 
for each wavelength from the best fit to the visibility amplitudes ($\chi^2_{red} = 2.32$) and differential visibility phases 
($\chi^2_{red} = 1.44$). For convenience, we also show the hypergiant and the possible four predicted positions of the pulsar. Figure \ref{fig:combined size} shows 
the resulting size of the extended wind component for each wavelength. The asymmetry in the wind size across wavelength still remains, as in the wind-only model. 
The wind sizes are slightly increased due to the smaller flux in the wind. The differential phases, on the other hand, are explained by having a compact extra component 
represented by the gas stream. 

For comparison, we also show in Figure \ref{fig:combined sky} (bottom) a stream model in the sky plane. The model follows \cite{Leahy08}, and 
assumes that at each time some mass is ejected from the hypergiant star's surface that intersects the line-of-centers of the binary. The stream is 
then formed by propagating each mass element, assuming that the radial velocity follows the CAK wind velocity law and the angular velocity is given
by conservation of angular momentum (the hypergiant is rotating). For the model shown, we simply assumed the relevant parameters from 
Table\ref{table:BP Cru Properties}, and that the pulsar is located at position "1" ($i = 60 \degr$;$\Omega=0 \degr$) at the time of observation. The calculation is performed in the binary plane and then projected to the sky plane, with the colors along the stream
representing the radial velocity. The stream shape is very sensitive to the assumed parameters, but it could be an explanation for asymmetric 
differential visibility phases along the emission line.

\section{Additional Data and Future Work} 

Here, we present additional spectral data that hint 
at the next steps in the study of BP Cru with optical interferometry. 

As alluded above, the emission lines in BP Cru may be formed from multiple, distinct components which are either not apparent at the moderate spectral resolution of GRAVITY ($R \sim 4,000$) or are modulated by the pulsar's radial velocity curve  ($v \sim 218 \text{ km/s}$), such as for an accretion disk or possibly a gas stream. This would complicate our model fitting from the previous section. 

For these reasons, we have compared the GRAVITY K band spectrum with that measured by XSHOOTER, 
using archival data\footnote{based on observations with ESO Telescopes at the La Silla Paranal Observatory under 
programme ID 095.C-0446(A)} reduced with the publicly available ESO XSHOOTER pipeline. 
It has a substantially higher spectral resolution ($R \sim 11,500$) than GRAVITY. 

\begin{figure}[tb]
\centering
\includegraphics[width=\columnwidth, clip]{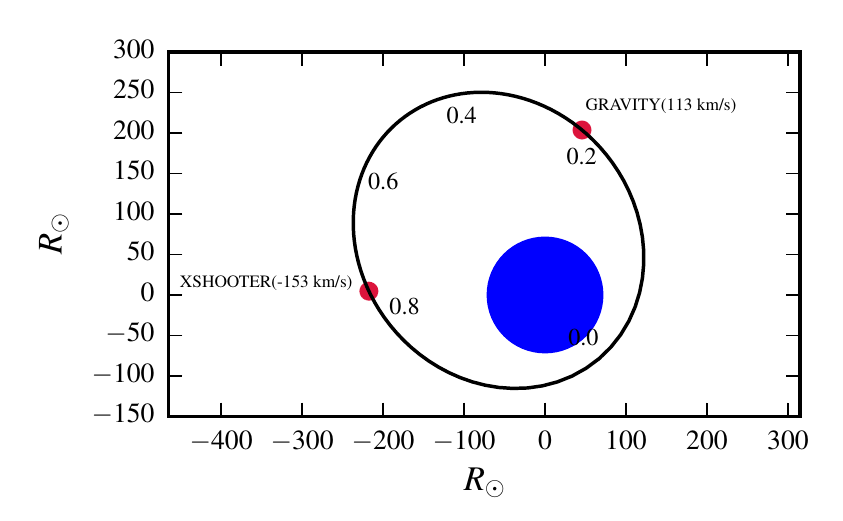} \\ 
\caption{Orbit of the pulsar in the binary plane. Orbital phases are indicated, as well as the positions at the time of observations and the corresponding radial velocity. The donor star is shown in blue with the photospheric radius $\sim 70 R_{\odot}$.}
\label{fig:orbit velocities}
\end{figure}

Figure \ref{fig:orbit velocities} shows the orbit of the pulsar in the binary plane, as well as the positions 
of the pulsar at the time of the GRAVITY and XSHOOTER observations. The radial velocities 
of the pulsar are also indicated. 

\begin{figure}[tb]
\centering
\includegraphics[width=\columnwidth, clip]{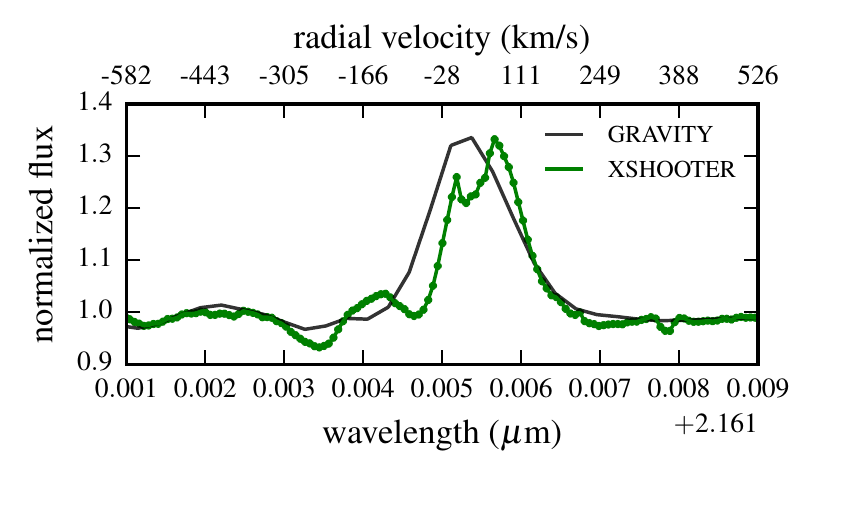} \\ 
\includegraphics[width=\columnwidth, clip]{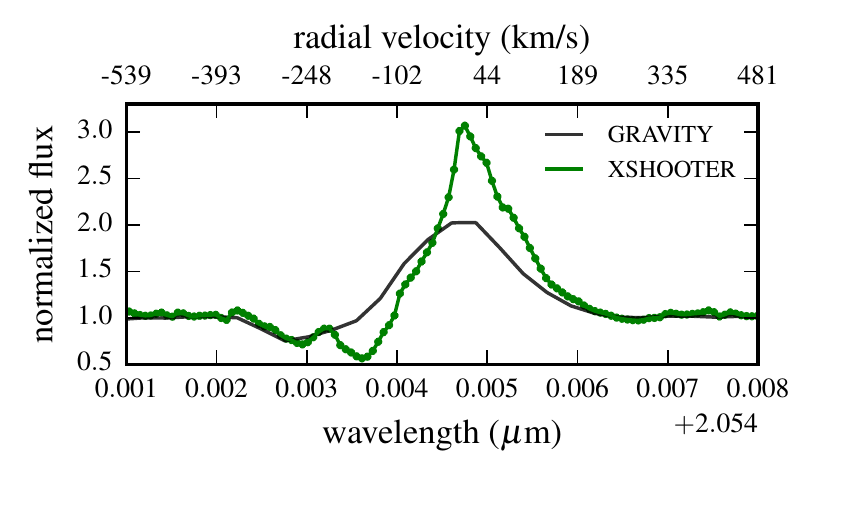} 
\caption{Br$\gamma$ and HeI $2.059 \mu$m lines as seen with GRAVITY UT and XSHOOTER at different orbital phases. The latter has a higher spectral resolution than the former ($R\sim11,500$ vs $4,000$) and shows substructure indicative of multiple line emission components.}
\label{fig:spectra instruments}
\end{figure}

Figure \ref{fig:spectra instruments} shows the spectra at the HeI $2.059 \mu$m and Br$\gamma$ emission lines for the two instruments. The higher resolution XSHOOTER spectra shows substructure that suggests a more complex line emission, possibly with multiple components. It could therefore be that the line emission has both a contribution from the normal hypergiant wind as well as from a dense gas stream, as is the case for the H$\alpha$ line in Cygnus X-1 \citep{Yan08}. We note, in particular, what appears to be a blueshifted ($\sim-130$ km/s) emission component with $\sim 15\%$ of the main line strength, when the predicted pulsar radial velocity at the XSHOOTER orbital phase is $-150 \text{ km/s}$. If they indeed trail the pulsar, such components would be redshifted at the time of the GRAVITY UT interferometric observation and could potentially be related to the interferometric signatures in the red part of the line. 

Additional high-resolution spectra at different orbital phases could confirm the presence of such emission components. 
When coupled with interferometric data, they would also be highly beneficial in 
testing the different models. Just to mention a few, a comparison between apastron and periastron epochs would 
help to assess X-ray effects, a comparison between superior and inferior conjunctions could probe the effects of 
the pulsar at different parts (red versus blue) of the wind and the wavelength at which the interferometric 
signatures peak could indicate, with the help of high resolution spectroscopy, the line emission component that 
is responsible for the interferometric signatures. All of these could help, for instance, in differentiating between 
an extended and distorted wind model from a gas stream model or possibly show the need for a combined 
model. 

Finally, we note that the possibility that the differential signatures reported here could be related to the intrinsic variability of the stellar wind of the hypergiant cannot be absolutely excluded with the present data. Differential visibility amplitude and phase signatures have been observed previously in the H$\alpha$ and Br$\gamma$ lines of Rigel, a late-B supergiant \citep{Chesneau10, Chesneau14}. In this case, however, the lines are in absorption and the extension of the wind emission in Br$\gamma$ is found to originate close to the photosphere ($\sim1.25 R_*$), in contrast to the case of BP Cru. Nevertheless, optical spectroscopy monitoring of the isolated early-B hypergiants mentioned in Section \ref{spectrum} has detected variability in the P Cygni-type profiles of wind-sensitive lines, in the form of discrete absorption components that could be associated with non-spherical density perturbations \citep{Rivinius97}. High spectral resolution interferometric observations of such stars would help to assess whether such variability could cause differential signatures of the same scale as what is seen in BP Cru, or whether the gravitational and radiation fields of the X-ray pulsar are indeed determinant. 

\section{Summary}

We have shown a first analysis of near-infrared interferometric data of the HMXB BP Cru obtained with VLTI/GRAVITY:

\begin{enumerate}
\item The spectrum shows unusual  Br$\gamma$ emission for a star of its spectral type; the higher mass-loss rate may be related to an intrinsically denser wind or, as has been proposed from the X-ray data on this source, to a gas stream of enhanced density; 
\item The continuum visibilities suggest a uniform stellar disk of radius $\sim 1 R_*$, compatible with the still low infrared excess due to the wind in the K band; 
\item Spectral differential interferometry shows differential visibility amplitudes and phases across the Br$\gamma$ and HeI $2.059 \mu$m emission lines; 
\item Any model for the emission lines must produce asymmetric, extended structure and a smooth spatial centroid gradient with radial velocity;
\item Examples of physically motivated, geometrical models satisfying these constraints include scenarios where the Br$\gamma$ is dominated by an extended ($R \simeq 4-7 R_*$), distorted wind or by a combination of extended wind and high density gas stream; 
\item Further orbital phase resolved high resolution spectroscopy and interferometric observations could help to distinguish between models.
\end{enumerate}

To our knowledge, this is the first dataset probing HMXB spatial
structure on such small microarcsecond scales, in which the
interaction between the donor star and the pulsar is expected to
occur. Follow up studies may offer the possibility of testing the
accretion mechanism and, more generally, the gravitational and
radiation effects of the compact object on the stellar environment in
these exotic systems. 

\section*{}

\acknowledgments

Based on observations made with ESO Telescopes at the
La Silla Paranal Observatory under program ID 60.A-9102. 
We thank the technical, administrative and scientific staff 
of the participating institutes and the ESO Paranal observatory for their extraordinary support
during the development, installation and commissioning of GRAVITY. This research has
made use of the Jean-Marie Mariotti Center \texttt{Aspro}, \texttt{OIFits Explorer} and
\texttt{SearchCal services}, and of CDS Astronomical Databases \texttt{SIMBAD} and
\texttt{VIZIER}.

\appendix

\section{Pulsar positions on the Sky Plane}
\label{app:A}

Here we estimate the predicted pulsar positions in the sky plane (centered on the donor star) 
at the time of observation based on what is currently known about the system . In addition to the 
orbital parameters determined from the pulsar's radial velocity curve \citep{Koh97}, the 
following parameters are in theory needed: 

\begin{enumerate}
\item The binary inclination $i$; 
\item The mass ratio $q$; 
\item The longitude of the ascending node $\Omega$; 
\end{enumerate} 

In practice $q$ is not important because the donor star is much more massive than the pulsar. 

We adopt the inclination $i = 60 \degr \text{ or } 120 \degr \pm 10 \degr$ from \cite{Kaper06}, which is estimated based on the upper limit on the neutron star mass and the absence of X-ray eclipsing. This allows to estimate $a_X \approx 0.28$ mas from $a_X \sin i$ known from the pulsar's radial velocity amplitude. From the mass ratio $q = \frac{M_X}{M_{opt}} \approx 0.046$ estimated in \cite{Kaper06} from Wray's radial velocity curve, we estimate $a_{opt} = q a_X \approx 0.01$ mas, and therefore the semi-major of the relative orbit $a_{rel} = a_X + a_{opt} \approx 0.29$ mas $= 191.7 R_{\odot}$. The only remaining parameter to determine is $\Omega$, of which radial velocity measurements are completely independent. However, we may constrain $\Omega$ from X-ray and column density measurements. \cite{Kaper06} claims that the pulsar is behind Wray 977 in the orbital phase interval $0.18 \lesssim \phi \lesssim 0.34$ based on the decrease in X-ray flux after periastron passage due to absorption by the dense stellar wind, as well as an increase in column density. This allows to estimate $\Omega$ by setting $x$, the pulsar position in the sky plane, to zero when $\phi \approx 0.21$: 

\begin{equation}
x \propto \cos\Omega\cos(\omega+\nu) - \sin\Omega\sin(\omega+\nu)\cos i
\end{equation} 

where $\nu$ is the true anomaly, which depends on $\phi$ and $e$ only. Plugging in the appropriate values, we get 

\begin{equation}
\tan \Omega \sim \cot(7.85) \cos i \Rightarrow \Omega \sim 0 \degr
\end{equation}

Therefore, there are four solutions for the pulsar position, corresponding to $(i,\Omega) \sim (60 \degr, 0 \degr)$, $(60 \degr, 180 \degr)$, $(120 \degr, 0 \degr)$, $(120 \degr, 180 \degr)$. They all have the same radial velocity solution and the same orbital phase at superior conjunction, and therefore cannot be distinguished with what is currently known about the system. 

Figure \ref{fig: orbit positions} shows the four possible positions of the pulsar on the sky plane (centered on Wray 977), along with the six baseline directions.  

\section{Correcting for the Photospheric Spectrum}
\label{app:B}

\begin{figure}[tb]
\centering
\includegraphics[width=\columnwidth/2, clip]{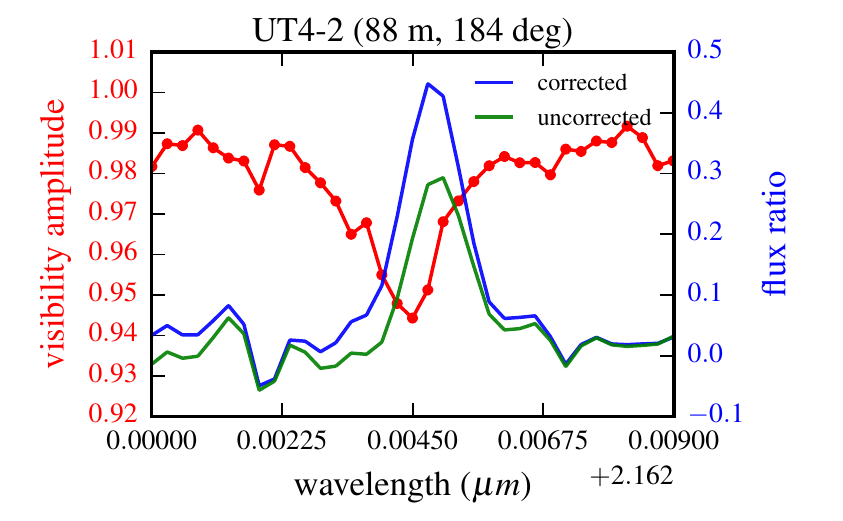} \\ 
\caption{Differential visibility amplitude (red) across the Br$\gamma$ line for one baseline (red), together with the flux ratio obtained from 
the spectrum assuming a flat continuum (green) and a continuum that contains a photospheric line (blue). The latter yields a flux ratio 
$\sim 50\%$ larger, as well as an increase in the blue portion of the line, in which the interferometric signatures are largest.}
\label{fig:flux correction}
\end{figure}

Figure \ref{fig:flux correction} shows the visibility amplitude on top of the flux ratio (blue) along the Br$\gamma$ region 
for baseline UT4-2, with the flux ratio taken directly from the spectrum by assuming a flat continuum (i.e. continuum $=1$ 
in the normalized spectrum). Especially on the blue side of the line, it is clear that interferometric signatures occur at regions 
where the flux ratio is near zero, which is confusing at first. However, one must remember that the unresolved part of the flux 
(i.e. the "continuum") includes photospheric absorption lines, which get filled by the emission component(s) in the combined 
spectrum. This is especially clear from the spectra of the comparison stars in Figure\ref{fig:spectra comparison}, which actually 
show absorption in Br$\gamma$, likely due to their $\sim 5-10 \times$ smaller mass-loss rate. 

Therefore, in order to obtain a more correct value for the flux ratio between the emission component(s) and the unresolved continuum, 
we must estimate the purely photospheric spectrum of Wray 977. One possibility would be to use stellar atmosphere model codes and 
set an artificially lower mass-loss rate. Since this is beyond the scope of this paper, we take a simpler approach and use the spectrum 
of an isolated blue supergiant star of the same spectral type to estimate the photospheric spectrum. Contrary to the H-band Brackett 
lines, the Br$\gamma$ line depth is not very sensitive to the star's luminosity/gravity \citep{Hanson96}; therefore, the spectrum of a 
smaller star, with a lower luminosity and much weaker wind, should be a good approximation to the spectrum of Wray's photosphere, 
at least at the Br$\gamma$ line. 

With this in mind, we chose the star HD 148688 (B1Ia), with K band spectrum available from \cite{Hanson05}. After degrading the original 
resolution ($R\sim12,000$) to GRAVITY's, we divide the GRAVITY spectrum by it, resulting in $1 + f$, where $f$ is the flux ratio between 
emission and photosphere. This "photospheric corrected" flux ratio is also shown in Figure \ref{fig:flux correction}. 

We note that such a correction ameliorates the presence of interferometric signatures at vanishing flux ratios, as the corrected flux 
ratio is shifted to the blue (an effect due to the wind emission being slightly redshifted with respect to the photosphere). Additionally, 
is should be more representative of the true flux ratio. Unfortunately, this method does not work for the HeI $2.059 \mu$m line, as it is 
very sensitive to winds and, unlike Br$\gamma$, goes easily into emission even for this star i.e. its photospheric spectrum is not 
easily recoverable. 

\section{The marginally resolved limit of inteferometry}
\label{app:C}

In general, differential visibility amplitudes and phases carry information about the source structure and are therefore model-dependent, which leads to difficulty in their interpretation if not enough uv-coverage is available or if the model is complicated or unknown. Here, however, we show that differential visibilities can, under certain conditions, provide robust, model-independent estimates about the image. The analysis is similar to that in \cite{Lachaume03}, but we focus in spectral differential quantities instead. 

From the Van Cittert-Zernike theorem, 
\begin{equation}
F(\vect{u}) = \iint I (\vect{\sigma}) e^{-2\pi i \vect{\sigma} \cdot \vect{u}} dl dm 
\label{eq:cittert}
\end{equation}
where $F$ is the coherent flux, $I$ is the source intensity distribution, $\vect{\sigma} = (l,m)$ are the object coordinates on sky and $\vect{u} = \frac{\vect{B}}{\lambda} = (u,v)$ is the baseline vector. 
In the following, it will be useful to define the moments of the intensity distribution about the origin as 
\begin{equation}
\mu_{pq} = \iint I(\vect{\sigma}) l^p m^q dl dm
\end{equation}
so that, for example, the zero-order moment $\mu_{00}$ is the total intensity and the normalized first-order moments $l_1 = \frac{\mu_{10}}{\mu_{00}}$ and $m_1 = \frac{\mu_{01}}{\mu_{00}}$ are the centroid positions along the $l$- and $m$-axes respectively. 
We can expand the complex exponential term in the integral of Eq.\ (\ref{eq:cittert}) in a Taylor series 
\begin{equation}
e^{-2\pi i \vect{\sigma} \cdot \vect{u}} = 1 - 2 \pi i (\vect{\sigma} \cdot \vect{u}) - 2 \pi^2 (\vect{\sigma} \cdot \vect{u})^2 + \frac{4\pi^3 i}{3} (\vect{\sigma} \cdot \vect{u})^3 + \mathcal{O} ((\vect{\sigma} \cdot \vect{u})^4)
\end{equation} 
which allows the use of approximations when 
\begin{equation}
|\vect{\sigma} \cdot \vect{u}| \ll 1\Leftarrow |\vect{\sigma}| \ll \frac{\lambda}{|\vect{B}|} 
\end{equation} 
i.e. when the source is sufficiently unresolved for a given baseline vector. 
Using the standard definition of the complex visibility
\begin{equation}
V(\vect{u}) = \frac{F(\vect{u})}{F(\vect{0})} = \frac{F(\vect{u})}{\mu_{00}}
\end{equation} 
it follows that 
\begin{equation}
V(\vect{u}) \approx 1 - 2 \pi i w_1 - 2 \pi^2 w_2 + \frac{4\pi^3 i}{3} w_3
\end{equation}
where 
\begin{align}
w_i = \frac{1}{\mu_{00}} \int  I(\vect{\sigma}) (\vect{\sigma} \cdot \vect{u})^i dl dm 
\label{eq:w} 
\end{align}
\subsection{Differential Visibility Phases as Centroid Probes} 
To first order in $\vect{\sigma} \cdot \vect{u}$, the phase of the visibility is 
\begin{equation}
arg(V(\vect{u})) \approx \arctan(\frac{-2\pi w_1}{1}) \approx -2 \pi w_1
\end{equation}
since $w_1 \ll 1$. Calling $\vect{x} = (l_1,m_1)$ the centroid positions for the given intensity distribution, 
\begin{equation}
arg(V(\vect{u})) \approx -2 \pi \vect{u} \cdot \vect{x} 
\label{eq:argV} 
\end{equation}
For given two images $a$ and $b$ at the same spatial frequency $\vect{u}$, the differential phase
\begin{align}
\Delta \phi_{ba} = arg(V_b) - arg(V_a) \nonumber \\ \approx -2 \pi (w_{1,b}-w_{1,a}) \nonumber \\ \approx -2 \pi \vect{u} \cdot (\vect{x}_b - \vect{x}_a) 
\label{eq:phase}
\end{align} 
Eq.\ (\ref{eq:phase}) shows that \textit{differential visibility phases give model-independent centroid displacements along the baseline direction for close to unresolved sources}. If two or more baselines are available, this allows to solve or fit for the centroid displacement $\vect{\Delta x}_{ab}$. 
Note that the differential phase is proportional to the baseline length. Therefore, for this approximation method to work in practice as a robust, model-independent estimation, we must have sufficiently small differential phase errors so that a signal can be measured even with a small enough baseline so that the sources remain very close to unresolved. Fortunately, this is exactly the case in spectral differential phase measurements, for which the error is much smaller then the absolute phase errors plagued by systematics.   
\subsection{Differential Visibility Amplitudes as Size/Asymmetry Probes} 
In case the source is close to but not completely unresolved, it is possible to obtain further robust, model-independent information about the image by using differential visibility amplitudes. To second-order in $\vect{\sigma} \cdot \vect{u}$, 
\begin{align}
|V(\vect{u})| \approx ((1-2\pi^2 w_2)^2 + (2\pi w_1)^2)^{1/2} \\ \approx 1 + 2 \pi^2 w_1^2 - 2 \pi^2 w_2 +2 \pi^4 w_2^2 \\ \approx 1 + 2 \pi^2 (w_1^2 -  w_2) 
\end{align}
since $w_1,  w_2 \ll 1$ and where we must expand to second order since the first-order term alone would result in $|V| > 1$. Note that in this expression the 
visibility amplitude depends on $w_1$ i.e. on the centroid of the image and therefore on the absolute phase, which is not available from single-axis 
interferometry. Even the differential visibility amplitude between two images $a$ and $b$ with this expression would depend on $w_{1,b}^2 - w_{1,a}^2$, 
whereas only $w_{1,b}-w_{1,a}$ is available from the differential visibility phase as shown above. In order to circumvent this, it is useful to define the 
moments of the image with respect to the centroid $\vect{x} = (l_1,m_1)$
\begin{equation}
\tilde{\mu}_{pq} = \iint I(\vect{\sigma}) (l-l_1)^p (m-m_1)^q dl dm
\end{equation}
so that, for example, the normalized second-order moments $\tilde{l}_2 = \frac{\tilde{\mu}_{20}}{\mu_{00}}$ and $\tilde{m}_2 = \frac{\tilde{\mu}_{02}}{\mu_{00}}$ are the variances about the centroid position along the $l$- and $m$-axes respectively, and $\frac{\tilde{\mu}_{11}}{\mu_{00}}$ is the covariance. Analogously, we define 
\begin{align}
\tilde{w_i} = \frac{1}{\mu_{00}} \int  I(\vect{\sigma}) ((\vect{\sigma}-\vect{x}) \cdot \vect{u})^i dl dm 
\label{eq:wtilde} 
\end{align}
It is straightforward to show directly from the definitions that $\tilde{w_2} = w_2 - w_1^2$, so that 
\begin{align}
|V| \approx 1 - 2\pi^2 \tilde{w_2} 
\end{align}
where by definition (Eq.(\ref{eq:wtilde})), for a given baseline $\vect{u}=(u,v)$
\begin{equation}
\tilde{w_2} = u^2 \frac{\tilde{\mu}_{20}}{\mu_{00}} + v^2 \frac{\tilde{\mu}_{02}}{\mu_{00}} + 2 uv \frac{\tilde{\mu}_{11}}{\mu_{00}}
\end{equation}
Note that this is a better definition since these moments are about the centroid of the origin rather than an arbitrary phase center. 
Given two images $a$ and $b$, for example at the continuum and at a spectral line, the differential visibility amplitude is therefore 
\begin{equation}
\Delta|V|_{ba} = |V|_b - |V|_a \approx -2\pi^2 (\tilde{w_{2,b}} - \tilde{w_{2,a}}) 
\label{eq:amplitude}
\end{equation}
Therefore, if three or more baselines are available, it is possible to solve for the difference in variances and covariance about the centroid 
between the continuum and the spectral line images. If a model for the continuum is available, \textit{differential visibility amplitudes allow 
obtaining robust estimates of the variances about the centroid position, which are related to the image size, as well as the covariance, 
which is related to the image asymmetry.}
\subsection{Closure Phases} 
Note that Eq.(\ref{eq:argV}) implies that, for any baseline triangle $\vect{u_1}+\vect{u_2}+\vect{u_3}=0$, the closure phase
\begin{align}
arg(V(\vect{u_1}))+arg(V(\vect{u_2}))+arg(V(\vect{u_3})) = -2 \pi (\vect{u_1}+\vect{u_2}+\vect{u_3}) \cdot \vect{x} = 0
\end{align}
Therefore, the close to unresolved limit must be compatible with vanishing closure phases for all baselines. Note also that vanishing closure phases do not necessarily imply a centro-symmetric structure, as these would have visibility phases of $0 \degr$ or $180 \degr$. 

Formally, this only happens because we have only 
kept the $\mathcal{O} (\vect{\sigma} \cdot \vect{u})$ term in the expansion. It can be shown \citep{Lachaume03} that the closure phases are related to the third-order moments of the image distribution, and therefore only contain terms starting at $\mathcal{O} ((\vect{\sigma} \cdot \vect{u})^3)$. Therefore, although the closure phases don't vanish absolutely, they are expected to be much smaller than the visibility phases themselves in the marginally resolved limit, and very likely cannot be detected within the noise limit of the instrument. 
\subsection{Validity of the Approximation} 
We have shown that the marginally resolved limit is applicable when $|\vect{\sigma} \cdot \vect{u}| \ll 1$. The translation of this condition into a minimum $|V|$, and the error incurred in the approximation, obviously depends on the baseline $\vect{u}$ and on the model itself. \cite{Lachaume03} compared the exact versus the approximated visibilities for different simple models (binary, ring, gaussian disc) and found that the approximation holds (i.e. the models are indistinguishable) up to $|V| \gtrsim 0.9$ (see their Figure 4).

\allauthors


\begin{thebibliography}{}
\expandafter\ifx\csname natexlab\endcsname\relax\def\natexlab#1{#1}\fi
\providecommand{\url}[1]{\href{#1}{#1}}

\bibitem[{{Bandyopadhyay} {et~al.}(1999){Bandyopadhyay}, {Shahbaz}, {Charles},
  \& {Naylor}}]{Bandyopadhyay99}
{Bandyopadhyay}, R.~M., {Shahbaz}, T., {Charles}, P.~A., \& {Naylor}, T. 1999,
  \mnras, 306, 417

\bibitem[{{Blondin}(1994)}]{Blondin94}
{Blondin}, J.~M. 1994, \apj, 435, 756

\bibitem[{{Bondi} \& {Hoyle}(1944)}]{bondihoyle1944}
{Bondi}, H., \& {Hoyle}, F. 1944, \mnras, 104, 273

\bibitem[{{Castor} {et~al.}(1975){Castor}, {Abbott}, \& {Klein}}]{Castor75}
{Castor}, J.~I., {Abbott}, D.~C., \& {Klein}, R.~I. 1975, \apj, 195, 157

\bibitem[{{Chesneau} {et~al.}(2014){Chesneau}, {Kaufer}, {Stahl}, {Colvinter},
  {Spang}, {Dessart}, {Prinja}, \& {Chini}}]{Chesneau14}
{Chesneau}, O., {Kaufer}, A., {Stahl}, O., {et~al.} 2014, \aap, 566, A125

\bibitem[{{Chesneau} {et~al.}(2010){Chesneau}, {Dessart}, {Mourard},
  {B{\'e}rio}, {Buil}, {Bonneau}, {Borges Fernandes}, {Clausse}, {Delaa},
  {Marcotto}, {Meilland}, {Millour}, {Nardetto}, {Perraut}, {Roussel}, {Spang},
  {Stee}, {Tallon-Bosc}, {McAlister}, {ten Brummelaar}, {Sturmann}, {Sturmann},
  {Turner}, {Farrington}, \& {Goldfinger}}]{Chesneau10}
{Chesneau}, O., {Dessart}, L., {Mourard}, D., {et~al.} 2010, \aap, 521, A5

\bibitem[{{Choquet} {et~al.}(2014){Choquet}, {Kervella}, {Le Bouquin},
  {M{\'e}rand}, {Berger}, {Haubois}, {Perrin}, {Petrucci}, {Lazareff}, \&
  {Pott}}]{Choquet14}
{Choquet}, {\'E}., {Kervella}, P., {Le Bouquin}, J.-B., {et~al.} 2014, \aap,
  561, A46

\bibitem[{{Clark} {et~al.}(2003){Clark}, {Charles}, {Clarkson}, \&
  {Coe}}]{Clark03}
{Clark}, J.~S., {Charles}, P.~A., {Clarkson}, W.~I., \& {Coe}, M.~J. 2003,
  \aap, 400, 655

\bibitem[{{Clark} {et~al.}(2012){Clark}, {Najarro}, {Negueruela}, {Ritchie},
  {Urbaneja}, \& {Howarth}}]{Clark12}
{Clark}, J.~S., {Najarro}, F., {Negueruela}, I., {et~al.} 2012, \aap, 541, A145

\bibitem[{{Eisenhauer} {et~al.}(2011){Eisenhauer}, {Perrin}, {Brandner},
  {Straubmeier}, {Perraut}, {Amorim}, {Sch{\"o}ller}, {Gillessen}, {Kervella},
  {Benisty}, {Araujo-Hauck}, {Jocou}, {Lima}, {Jakob}, {Haug}, {Cl{\'e}net},
  {Henning}, {Eckart}, {Berger}, {Garcia}, {Abuter}, {Kellner}, {Paumard},
  {Hippler}, {Fischer}, {Moulin}, {Villate}, {Avila}, {Gr{\"a}ter}, {Lacour},
  {Huber}, {Wiest}, {Nolot}, {Carvas}, {Dorn}, {Pfuhl}, {Gendron}, {Kendrew},
  {Yazici}, {Anton}, {Jung}, {Thiel}, {Choquet}, {Klein}, {Teixeira}, {Gitton},
  {Moch}, {Vincent}, {Kudryavtseva}, {Str{\"o}bele}, {Sturm}, {F{\'e}dou},
  {Lenzen}, {Jolley}, {Kister}, {Lapeyr{\`e}re}, {Naranjo}, {Lucuix},
  {Hofmann}, {Chapron}, {Neumann}, {Mehrgan}, {Hans}, {Rousset}, {Ramos},
  {Suarez}, {Lederer}, {Reess}, {Rohloff}, {Haguenauer}, {Bartko}, {Sevin},
  {Wagner}, {Lizon}, {Rabien}, {Collin}, {Finger}, {Davies}, {Rouan},
  {Wittkowski}, {Dodds-Eden}, {Ziegler}, {Cassaing}, {Bonnet}, {Casali},
  {Genzel}, \& {Lena}}]{Eisenhauer11}
{Eisenhauer}, F., {Perrin}, G., {Brandner}, W., {et~al.} 2011, The Messenger,
  143, 16

\bibitem[{{Evangelista} {et~al.}(2010){Evangelista}, {Feroci}, {Costa}, {Del
  Monte}, {Donnarumma}, {Lapshov}, {Lazzarotto}, {Pacciani}, {Rapisarda},
  {Soffitta}, {Argan}, {Barbiellini}, {Boffelli}, {Bulgarelli}, {Caraveo},
  {Cattaneo}, {Chen}, {D'Ammando}, {Di Cocco}, {Fuschino}, {Galli}, {Gianotti},
  {Giuliani}, {Labanti}, {Lipari}, {Longo}, {Marisaldi}, {Mereghetti},
  {Moretti}, {Morselli}, {Pellizzoni}, {Perotti}, {Piano}, {Picozza}, {Pilia},
  {Prest}, {Pucella}, {Rappoldi}, {Sabatini}, {Striani}, {Tavani}, {Trifoglio},
  {Trois}, {Vallazza}, {Vercellone}, {Vittorini}, {Zambra}, {Antonelli},
  {Cutini}, {Pittori}, {Preger}, {Santolamazza}, {Verrecchia}, {Giommi}, \&
  {Salotti}}]{Evangelista10}
{Evangelista}, Y., {Feroci}, M., {Costa}, E., {et~al.} 2010, \apj, 708, 1663

\bibitem[{{Foreman-Mackey} {et~al.}(2013){Foreman-Mackey}, {Hogg}, {Lang}, \&
  {Goodman}}]{Foreman13}
{Foreman-Mackey}, D., {Hogg}, D.~W., {Lang}, D., \& {Goodman}, J. 2013, \pasp,
  125, 306

\bibitem[{{Fuerst} {et~al.}(2016){Fuerst}, {Kreykenbohm}, {Kretschmar},
  {Ballhausen}, \& {Pottschmidt}}]{Fuerst16}
{Fuerst}, F., {Kreykenbohm}, I., {Kretschmar}, P., {Ballhausen}, R., \&
  {Pottschmidt}, K. 2016, The Astronomer's Telegram, 8870

\bibitem[{{Haberl}(1991)}]{Haberl91}
{Haberl}, F. 1991, \apj, 376, 245

\bibitem[{{Hanson} {et~al.}(1996){Hanson}, {Conti}, \& {Rieke}}]{Hanson96}
{Hanson}, M.~M., {Conti}, P.~S., \& {Rieke}, M.~J. 1996, \apjs, 107, 281

\bibitem[{{Hanson} {et~al.}(2005){Hanson}, {Kudritzki}, {Kenworthy}, {Puls}, \&
  {Tokunaga}}]{Hanson05}
{Hanson}, M.~M., {Kudritzki}, R.-P., {Kenworthy}, M.~A., {Puls}, J., \&
  {Tokunaga}, A.~T. 2005, \apjs, 161, 154

\bibitem[{{Islam} \& {Paul}(2014)}]{Islam14}
{Islam}, N., \& {Paul}, B. 2014, \mnras, 441, 2539

\bibitem[{{Kaper} {et~al.}(2006){Kaper}, {van der Meer}, \&
  {Najarro}}]{Kaper06}
{Kaper}, L., {van der Meer}, A., \& {Najarro}, F. 2006, \aap, 457, 595

\bibitem[{{Koh} {et~al.}(1997){Koh}, {Bildsten}, {Chakrabarty}, {Nelson},
  {Prince}, {Vaughan}, {Finger}, {Wilson}, \& {Rubin}}]{Koh97}
{Koh}, D.~T., {Bildsten}, L., {Chakrabarty}, D., {et~al.} 1997, \apj, 479, 933

\bibitem[{{Kreykenbohm} {et~al.}(2004){Kreykenbohm}, {Wilms}, {Coburn},
  {Kuster}, {Rothschild}, {Heindl}, {Kretschmar}, \&
  {Staubert}}]{Kreykenbohm04}
{Kreykenbohm}, I., {Wilms}, J., {Coburn}, W., {et~al.} 2004, \aap, 427, 975

\bibitem[{{Kudritzki} \& {Puls}(2000)}]{Kudritzki00}
{Kudritzki}, R.-P., \& {Puls}, J. 2000, \araa, 38, 613

\bibitem[{{Lachaume}(2003)}]{Lachaume03}
{Lachaume}, R. 2003, \aap, 400, 795

\bibitem[{{Lapeyrere} {et~al.}(2014){Lapeyrere}, {Kervella}, {Lacour},
  {Azouaoui}, {Garcia-Dabo}, {Perrin}, {Eisenhauer}, {Perraut}, {Straubmeier},
  {Amorim}, \& {Brandner}}]{Lapeyrere14}
{Lapeyrere}, V., {Kervella}, P., {Lacour}, S., {et~al.} 2014, in \procspie,
  Vol. 9146, Optical and Infrared Interferometry IV, 91462D

\bibitem[{{Leahy}(1991)}]{Leahy91}
{Leahy}, D.~A. 1991, \mnras, 250, 310

\bibitem[{{Leahy}(2002)}]{Leahy02}
---. 2002, \aap, 391, 219

\bibitem[{{Leahy} \& {Kostka}(2008)}]{Leahy08}
{Leahy}, D.~A., \& {Kostka}, M. 2008, \mnras, 384, 747

\bibitem[{{Liu} {et~al.}(2006){Liu}, {van Paradijs}, \& {van den
  Heuvel}}]{Liu06}
{Liu}, Q.~Z., {van Paradijs}, J., \& {van den Heuvel}, E.~P.~J. 2006, \aap,
  455, 1165

\bibitem[{{Meilland} {et~al.}(2012){Meilland}, {Millour}, {Kanaan}, {Stee},
  {Petrov}, {Hofmann}, {Natta}, \& {Perraut}}]{Meilland12}
{Meilland}, A., {Millour}, F., {Kanaan}, S., {et~al.} 2012, \aap, 538, A110

\bibitem[{{Merand} {et~al.}(2005){Merand}, {Borde}, \& {Coud{\'e} du
  Foresto}}]{Merand05}
{Merand}, A., {Borde}, P., \& {Coud{\'e} du Foresto}, V. 2005, VizieR Online
  Data Catalog, 343

\bibitem[{{Perez M.} \& {Blundell}(2009)}]{Perez09}
{Perez M.}, S., \& {Blundell}, K.~M. 2009, \mnras, 397, 849

\bibitem[{{Pestalozzi} {et~al.}(2009){Pestalozzi}, {Torkelsson}, {Hobbs}, \&
  {L{\'o}pez-S{\'a}nchez}}]{Pestalozzi09}
{Pestalozzi}, M., {Torkelsson}, U., {Hobbs}, G., \& {L{\'o}pez-S{\'a}nchez},
  {\'A}.~R. 2009, \aap, 506, L21

\bibitem[{{Pravdo} {et~al.}(1995){Pravdo}, {Day}, {Angelini}, {Harmon},
  {Yoshida}, \& {Saraswat}}]{Pravdo95}
{Pravdo}, S.~H., {Day}, C.~S.~R., {Angelini}, L., {et~al.} 1995, \apj, 454, 872

\bibitem[{{Rivinius} {et~al.}(1997){Rivinius}, {Stahl}, {Wolf}, {Kaufer},
  {Gaeng}, {Gummersbach}, {Jankovics}, {Kovacs}, {Mandel}, {Peitz}, {Szeifert},
  \& {Lamers}}]{Rivinius97}
{Rivinius}, T., {Stahl}, O., {Wolf}, B., {et~al.} 1997, \aap, 318, 819

\bibitem[{{Servillat} {et~al.}(2014){Servillat}, {Coleiro}, {Chaty}, {Rahoui},
  \& {Zurita Heras}}]{Servillat14}
{Servillat}, M., {Coleiro}, A., {Chaty}, S., {Rahoui}, F., \& {Zurita Heras},
  J.~A. 2014, \apj, 797, 114

\bibitem[{{Shahbaz} {et~al.}(1999){Shahbaz}, {Bandyopadhyay}, \&
  {Charles}}]{Shahbaz09}
{Shahbaz}, T., {Bandyopadhyay}, R.~M., \& {Charles}, P.~A. 1999, \aap, 346, 82

\bibitem[{{Stevens}(1988)}]{Stevens88}
{Stevens}, I.~R. 1988, \mnras, 232, 199

\bibitem[{{Thureau} {et~al.}(2009){Thureau}, {Monnier}, {Traub},
  {Millan-Gabet}, {Pedretti}, {Berger}, {Garcia}, {Schloerb}, \&
  {Tannirkulam}}]{Thureau09}
{Thureau}, N.~D., {Monnier}, J.~D., {Traub}, W.~A., {et~al.} 2009, \mnras, 398,
  1309

\bibitem[{{{\v C}echura} \& {Hadrava}(2015)}]{Cechura15}
{{\v C}echura}, J., \& {Hadrava}, P. 2015, \aap, 575, A5

\bibitem[{{Walder} {et~al.}(2014){Walder}, {Melzani}, {Folini},
  {Winisdoerffer}, \& {Favre}}]{Walder14}
{Walder}, R., {Melzani}, M., {Folini}, D., {Winisdoerffer}, C., \& {Favre},
  J.~M. 2014, in Astronomical Society of the Pacific Conference Series, Vol.
  488, 8th International Conference of Numerical Modeling of Space Plasma Flows
  (ASTRONUM 2013), ed. N.~V. {Pogorelov}, E.~{Audit}, \& G.~P. {Zank}, 141

\bibitem[{{Walter} {et~al.}(2015){Walter}, {Lutovinov}, {Bozzo}, \&
  {Tsygankov}}]{Walter15}
{Walter}, R., {Lutovinov}, A.~A., {Bozzo}, E., \& {Tsygankov}, S.~S. 2015,
  \aapr, 23, 2

\bibitem[{{Yan} {et~al.}(2008){Yan}, {Liu}, \& {Hadrava}}]{Yan08}
{Yan}, J., {Liu}, Q., \& {Hadrava}, P. 2008, \aj, 136, 631

\end{thebibliography}
\end{document}